\documentclass{aa}
\usepackage[nolist]{acronym}
\usepackage[displaymath, mathlines]{lineno}

\usepackage{graphicx}
\usepackage{txfonts}
\usepackage{nameref}
\usepackage{nccmath}
\usepackage{subcaption}
\usepackage{siunitx}
\usepackage{color}
\usepackage{natbib,twoopt}
\bibpunct{(}{)}{;}{a}{}{,} 

\usepackage[hyphenbreaks]{breakurl}
\usepackage[breaklinks]{hyperref} 
\usepackage{cleveref}
\usepackage{scalerel}
\usepackage{tikz}
\usetikzlibrary{svg.path}

\definecolor{orcidlogocol}{HTML}{A6CE39}
\tikzset{
  orcidlogo/.pic={
    \fill[orcidlogocol] svg{M256,128c0,70.7-57.3,128-128,128C57.3,256,0,198.7,0,128C0,57.3,57.3,0,128,0C198.7,0,256,57.3,256,128z};
    \fill[white] svg{M86.3,186.2H70.9V79.1h15.4v48.4V186.2z}
                 svg{M108.9,79.1h41.6c39.6,0,57,28.3,57,53.6c0,27.5-21.5,53.6-56.8,53.6h-41.8V79.1z M124.3,172.4h24.5c34.9,0,42.9-26.5,42.9-39.7c0-21.5-13.7-39.7-43.7-39.7h-23.7V172.4z}
                 svg{M88.7,56.8c0,5.5-4.5,10.1-10.1,10.1c-5.6,0-10.1-4.6-10.1-10.1c0-5.6,4.5-10.1,10.1-10.1C84.2,46.7,88.7,51.3,88.7,56.8z};
  }
}

\newcommand\orcidicon[1]{\href{https://orcid.org/#1}{\mbox{\scalerel*{
\begin{tikzpicture}[yscale=-1,transform shape]
\pic{orcidlogo};
\end{tikzpicture}
}{|}}}}

\newcommand{\myorcidid}{0000-0001-9282-9462}

\definecolor{cobalt}{rgb}{0.06, 0.2, 0.65}
\hypersetup{
  colorlinks,
  citecolor=cobalt,
  linkcolor=[rgb]{0.8, 0.2, 1.0},
  urlcolor=cobalt,
}
\makeatletter

\renewcommand*\aa@pageof{, page \thepage{} of \pageref*{LastPage}}

\newcommand{\logg}{\ensuremath{\log g}}

\def\kms{\ifmmode{\rm km\th s^{-1}}\else km\th s$^{-1}$\fi}
\def\th{\thinspace}
\newcommand{\Mjup}{M$_{\text{Jup}}$}
\newcommand{\Rjup}{R$_{\text{Jup}}$}

\newcommand{\twelveCO}{\textsuperscript{12}CO}
\newcommand{\thirteenCO}{\textsuperscript{13}CO}

\newcommand{\eighteenO}{\textsuperscript{18}O}

\newcommand{\ratio}[3]{\textsuperscript{#2}#1/\textsuperscript{#3}#1}
\newcommand{\Htwo}{H$_2$}

\newcommand{\micron}{$\mu$m}
\newcommand{\vsini}{$v\sin{i}$\ }
\newcommand{\Teff}{$T_{\text{eff}}$}
\newcommand{\evidence}{$\mathcal{Z}$\ }
\newcommand{\Cratio}{\textsuperscript{12}C/\textsuperscript{13}C}
\newcommand{\Cratiosolar}{(\textsuperscript{12}C/\textsuperscript{13}C)$_{\odot}$}
\newcommand{\Cratioism}{(\textsuperscript{12}C/\textsuperscript{13}C)$_{\rm ISM}$}

\newcommand{\water}{H$_2$O\ }
\newcommand{\CRIRES}{CRIRES$^+$}
\newcommand{\sixteenOwater}{H\textsubscript{2}\textsuperscript{16}O}
\newcommand{\eighteenOwater}{H\textsubscript{2}\textsuperscript{18}O}

\newcommandtwoopt{\citeads}[3][][]{\href{http://adsabs.harvard.edu/abs/#3}%
{\def\hyper@linkstart##1##2{}%
    \let\hyper@linkend\@empty\citealp[#1][#2]{#3}}}
\newcommandtwoopt{\citepads}[3][][]{\href{http://adsabs.harvard.edu/abs/#3}%
{\def\hyper@linkstart##1##2{}%
    \let\hyper@linkend\@empty\citep[#1][#2]{#3}}}

\makeatother
\begin{document} 


\title{The $^{12}$C/$^{13}$C ratios of three young brown dwarfs with CRIRES$^+$}
\title{The ESO SupJup Survey \\ II. The $^{12}$C/$^{13}$C isotope ratios of three young brown dwarfs with CRIRES$^+$}

\titlerunning{The 12C/13C ratio of three young brown dwarfs with CRIRES$^+$}
\author{D. Gonz\'{a}lez Picos\inst{\ref{inst1} \orcidicon{\myorcidid}}
          \and
          I.A.G. Snellen\inst{\ref{inst1}} \and 
          S. de Regt\inst{\ref{inst1}} \and
          R. Landman\inst{\ref{inst1}} \and
          Y. Zhang\inst{\ref{inst2}, \ref{inst1}}\and
          S. Gandhi\inst{\ref{inst3},\ref{inst4},\ref{inst1}} \and C. Ginski\inst{\ref{inst5}} \and A. Y. Kesseli\inst{\ref{inst6}} \and P. Molli\`ere\inst{\ref{inst7}} \and
        T. Stolker\inst{\ref{inst1}}
}
\institute{
    Leiden Observatory, Leiden University, P.O. Box 9513, 2300 RA, Leiden, The Netherlands\\\email{picos@strw.leidenuniv.nl}\label{inst1} \and Department of Astronomy, California Institute of Technology, Pasadena, CA 91125, USA \label{inst2} \and
    Department of Physics, University of Warwick, Coventry CV4 7AL, UK\label{inst3} \and
    Centre for Exoplanets and Habitability, University of Warwick, Gibbet Hill Road, Coventry CV4 7AL, UK\label{inst4} \and
    School of Natural Sciences, Center for Astronomy, University of Galway, Galway, H91 CF50, Ireland\label{inst5} \and IPAC, Mail Code 100-22, Caltech, 1200 E. California Boulevard, Pasadena, CA 91125, USA \label{inst6} \and
    Max-Planck-Institut für Astronomie, Königstuhl 17, 69117 Heidelberg, Germany \label{inst7}
}

   \date{Received 19 March 2024 / Accepted 6 July 2024}

\abstract
{Young brown dwarfs exhibit atmospheric characteristics similar to those of super-Jupiters, providing a unique opportunity to study planetary atmospheres. Atmospheric retrievals of high-resolution spectra reveal detailed properties of these objects, with elemental and isotopic ratios offering insights into their formation history. The ESO SupJup Survey, utilising CRIRES$^+$ on the Very Large Telescope, aims to assess the role of \Cratio\ as a formation tracer.} 
{We present observations of three young brown dwarfs: 2MASS J12003792-7845082, TWA 28, and 2MASS J08561384-1342242. Our goal is to constrain their chemical compositions, thermal profiles, surface gravities, spin rotations, and \Cratio.} 
{We conducted atmospheric retrievals of CRIRES$^+$ \textit{K}-band spectra, coupling the radiative transfer code \texttt{petitRADTRANS} with the Bayesian inference algorithm \texttt{MultiNest}.} 
{The retrievals provide a detailed characterisation of the atmospheres of the three objects. We report the volume mixing ratios of the main molecular and atomic species: \sixteenOwater, \twelveCO, HF, Na, Ca, and Ti, including the novel detection of hydrogen fluoride (HF) in the atmosphere of a brown dwarf. We determine \Cratio\ values of $81^{+28}_{-19}$ and $79^{+20}_{-14}$ in the atmospheres of TWA 28 and J0856, respectively, with strong significance ($>3\sigma$). We also report tentative evidence ($\sim 2\sigma$) of \thirteenCO{} in J1200, at \Cratio = $114^{+69}_{-33}$. Additionally, we detect \eighteenOwater{} at moderate significance in J0856 (3.3$\sigma$) and TWA 28 (2.1$\sigma$). The retrieved thermal profiles are consistent with hot atmospheres ($2300-2600$ K) with low surface gravities and slow spins, as expected for young objects.}
{The measured carbon isotope ratios are consistent among the three objects and show no significant deviation from that of the local interstellar medium, suggesting a fragmentation-based formation mechanism similar to star formation. The tentative detection of \eighteenOwater{} in two objects of our sample highlights the potential of high-resolution spectroscopy to probe additional isotope ratios, such as \ratio{O}{16}{18}, in the atmospheres of brown dwarfs and super-Jupiters.}


\maketitle

\section{Introduction}

Young brown dwarfs (BDs) and gas giant exoplanets share atmospheric and spectral characteristics, making BDs excellent analogues for studying exoplanet atmospheres \citep{chauvinGiantPlanetCompanion2005, currieDIRECTIMAGINGSPECTROSCOPY2013, fahertyPOPULATIONPROPERTIESBROWN2016}. These young, self-luminous objects, characterised by high effective temperatures and inflated radii, serve as benchmarks for understanding the complex processes shaping planetary atmospheres \citep{marleyCoolSideModeling2015, madhusudhanExoplanetaryAtmospheresKey2019}. The advent of high-resolution spectroscopy (HRS) has transformed our ability to analyse these atmospheres, revealing details about their chemical compositions, thermal structures, and fundamental properties, such as rotational velocities and surface gravities \citep{snellenFastSpinYoung2014a, ruffioDeepExplorationPlanets2021}. 

Recent advancements in atmospheric retrievals have enabled the quantitative characterisation of the atmospheres of young BDs and directly imaged exoplanets \citep{molliereRetrievingScatteringClouds2020a, zhang13COrichAtmosphereYoung2021, wangRetrievingAbundancesHR2022, zhangELementalAbundancesPlanets2023a, landmanPictorisEyesUpgraded2024, inglisAtmosphericRetrievalsYoung2024}. Observations from state-of-the-art ground-based instruments like CRIRES$^+$ \citep{dornCRIRESSkyESO2023} and KPIC \citep{delormeKeckPlanetImager2021}, along with space-based facilities such as JWST \citep{rigbySciencePerformanceJWST2023}, have required more sophisticated models and retrieval frameworks to incorporate all physical and chemical processes in the atmosphere, in addition to the correct treatment of instrumental effects (e.g. \citealt{ruffioJWSTTSTHighContrast2023,deregtESOSupJupSurvey2024a}).

Knowing elemental abundances, such as the carbon-to-oxygen ratio (C/O), allows us to trace the formation mechanisms and origins of exoplanets \citep{obergEFFECTSSNOWLINESPLANETARY2011a}. The local conditions within a protoplanetary disk determine the material available for planet formation and the chemical composition of the resulting objects. As the temperature decreases with distance from the star, volatile species such as water and CO freeze out, leading to chemical segregation within the disk \citep{leeCarbonIsotopeChemistry2024}. Consequently, the C/O in resulting planets is linked to their birth location within the disk. Previous studies have shown distinct C/Os in exoplanets, ranging from C/O $\approx 0.1$ to C/Os close to unity for planets formed beyond the snow lines of water and CO (e.g. \citealt{obergEFFECTSSNOWLINESPLANETARY2011a,madhusudhanAtmosphericSignaturesGiant2017,brewerRATIOSSUGGESTHOT2017,lineSolarSubsolarMetallicity2021}). Similarly, measurements of nearby low-mass stars and BDs indicate C/Os varying from $\approx 0.3$ to $\approx 0.8$ \citep{nissenCarbontooxygenRatioStars2013,brewerMgSiRATIOS2016}. Precise measurements of C/O in stars are challenging due to the lack of molecular features in their spectra; the C/Os are typically inferred from the elemental abundances of C and O. Cooler objects, such as BDs and exoplanets, exhibit rich molecular spectra, allowing for a direct measurement of the C/O from molecular features in their spectra \citep{nissenCarbontooxygenRatioStars2013}. However, challenges arise for ultracool dwarfs due to the limitations of existing atmospheric model grids in fitting their spectra and the degeneracies between the C/O and other parameters such as the temperature profile and the presence of clouds \citep{phillipsCarbontooxygenRatioCool2024}.

Isotopic ratios, altered by chemical reactions and fractionation, may link the observed composition of objects to their formation environments. Objects that accrete material from the disk inherit its chemical composition, including isotopic ratios, potentially providing a link between the formation environment and the observed composition of the object \citep{molliereInterpretingAtmosphericComposition2022}. Isolated BDs are thought to form through molecular cloud collapse and fragmentation, akin to star formation processes \citep{bateFormationMechanismBrown2002}, or via disk fragmentation and ejection from extended disks around Sun-like stars \citep{stamatellosPropertiesBrownDwarfs2009}. In contrast, massive gas giant planets, with masses ranging from 5 to 30 \Mjup, hereafter referred to as super-Jupiters (SJs), are believed to arise from combined processes of solid core accretion and subsequent gas capture \citep{pollackFormationGiantPlanets1996} or disk fragmentation \citep{mayerFormationGiantPlanets2002}. Due to the uniqueness of these origins, chemical markers, such as the C/O and carbon isotopic ratios, may serve as indicators of their formation pathways. Precise measurements of these ratios in BDs and SJs are pivotal for understanding their formation mechanisms, offering insights into the broader phenomena of substellar and planetary formation \citep{nielsenGeminiPlanetImager2019, viganSPHEREInfraredSurvey2021, molliereInterpretingAtmosphericComposition2022, zhang13COrichAtmosphereYoung2021}.

In the Solar System, isotopic ratios serve as indispensable probes into the formation and evolutionary history of planets and smaller celestial bodies. Among these, the deuterium-to-hydrogen ratio (D/H) is particularly significant for unravelling the origins and historical dynamics of water \citep{aleonDeterminationInitialHydrogen2022}. This ratio is key to understanding various processes, including atmospheric loss and the potential roles of comets and asteroids in delivering water to planetary surfaces. The D/H isotope ratio varies across the Solar System, with a clear decreasing trend as a function of orbital distance: D/H $\sim \num{1.6e-2}$ on Venus \citep{donahueVenusWasWet1982}, $\sim \num{1.5e-4}$ on Earth \citep{hagemannAbsoluteIsotopicScale1970}, $\sim \num{2.0e-5}$ on the giant planets \citep{pierelRatiosSaturnJupiter2017}, and $\sim \num{4.0e-5}$ on the ice giants \citep{feuchtgruberRatioAtmospheresUranus2013}. Such measurements are crucial for piecing together the early environmental conditions and the subsequent development of habitable conditions on Earth-like planets \citep{goderisAncientImpactorComponents2016}.

The carbon isotope ratio across the Solar System is mostly uniform, \Cratiosolar$\sim 89$, with a terrestrial value of $89.3\pm0.2$ \citep{meijaIsotopicCompositionsElements2016} and a solar value of $93.5 \pm 3.1$ \citep{lyonsLightCarbonIsotope2018}. In contrast, the local present-day interstellar medium (ISM) exhibits a lower value of \Cratioism $=68\pm15$ \citep{milam1213Isotope2005}. The \Cratio\ is a potential tracer of formation pathways, as it is altered by chemical reactions and fractionation processes during the formation stages \citep{molliereInterpretingAtmosphericComposition2022}. Additional isotope ratios, such as \ratio{O}{16}{18}, \ratio{O}{16}{17}, and \ratio{N}{14}{15}, can provide complementary information and might be within the reach of current facilities \citep{gandhiJWSTMeasurements13C2023a,molliereInterpretingAtmosphericComposition2022, barrado15NH3AtmosphereCool2023}.

The finding of a \thirteenCO-rich atmosphere on the giant planet YSES 1b \citep{zhang13COrichAtmosphereYoung2021} sparked renewed interest in the \Cratio\ as a formation tracer. A subsequent analysis of a young, isolated BD revealed a high \Cratio\ \citep{zhang12CO132021}, higher than that of the ISM and comparable to the solar value. Additional observations with CRIRES confirmed a high \Cratio\ of $108\pm10$ in the young BD 2M0355 and hinted at the presence of C$^{18}$O \citep{zhangVLTCRIRESScience2022}. Recent HRS from Keck/KPIC has revealed solar-like values of \Cratio\ in the HIP 55507 system and in the hot BD HD 984 B \citep{xuanValidationElementalIsotopic2023,costesFreshViewHot2024}. Space-based observations have also provided valuable measurements of the \Cratio\ in the young, directly imaged planet VHS 1256 b \citep{gandhiJWSTMeasurements13C2023a} and a cool T dwarf \citep{hoodHighPrecisionAtmosphericConstraints2024}. In addition to the \Cratio, the \ratio{O}{16}{18} has been tentatively constrained in the HIP 55507, pointing to homogeneous values in the system \citep{xuanValidationElementalIsotopic2023}. An analysis of JWST/NIRSpec M-band spectra (4.1-5.3\micron{}) of the young SJ VHS 1256 b resulted in a \Cratio\ of $62 \pm 2$, in agreement with the value in the local ISM, within the uncertainties; the same analysis also found  additional isotope ratios for oxygen, including \ratio{O}{16}{18} and \ratio{O}{16}{17}, that were slightly lower than the ISM value
 \citep{gandhiJWSTMeasurements13C2023a}.

With the goal of testing the \Cratio\ as a formation tracer, the ESO SupJup Survey (PI: Snellen) observed a large sample of isolated BDs and SJs with CRIRES$^+$. An overview of the SupJup Survey is presented in \cite{deregtESOSupJupSurvey2024a}, along with the initial results, which focused on the late-L dwarf DENIS J0255-4700. In this work we present the results of the atmospheric retrievals of three young BDs from the SupJup Survey, including precise chemical abundances of the main molecular and atomic spectrally active species, tight constraints on thermal profiles, and derived C/O and \Cratio\ for each object. 

Section 2 provides an overview of the sample of young BDs. Next, we describe the observations and data reduction (see Sect. \ref{sec:observations}). In Sects. \ref{sec:forward_modeling} and \ref{sec:retrieval_framework} we present the forward modelling and retrieval framework. In Sect. \ref{sec:results} we report the results of the retrievals and discuss the implications of our findings in the context of young BDs and recent literature on isotope ratios. Finally, we summarise our conclusions in Sect. \ref{sec:conclusions}, highlighting the importance of HRS in characterising the atmospheres of young BDs and SJs and the potential of isotope ratios as tracers of their formation history.

\section{Sample of young isolated brown dwarfs}\label{sec:sample}
\begin{table*}
    \caption{Fundamental parameters of the young BDs.}
    \centering
    \renewcommand{\arraystretch}{1.5}
    \begin{tabular}{lccc}
        \hline
        \hline
         & 2MASS J12003792-7845082 & TWA 28 & 2MASS J08561384-1342242 \\
        \hline
        Distance (pc) & 101.1 $\pm$ 0.7$^{(1)}$ & 59.2 $\pm$ 0.4$^{(1)}$ & 53.8 $\pm$ 0.4$^{(1)}$ \\
        Age (Myr) & 4$^{(2)}$ & 5-10$^{(3)}$ & 10$^{(6)}$ \\
        SpT & M6$\gamma$$^{(2)}$ & M8.5$\gamma^{(4)}$ & M8.6$\beta^{(4)}$ \\
        T$_{\text{eff}}$ (K) & 2784-2850$^{(2)}$ & 2382$\pm$42$^{(4)}$ & 2380$\pm$32$^{(4)}$ \\
        Mass (M$_{\text{Jup}}$) & 42-58$^{(2)}$ & $20.9\pm$5$^{(3)}$ & $14.4\pm1.4^{(6)}$ \\
        \hline
    \end{tabular}\\
    \vspace{0.12cm}
{\bf References.} (1) \cite{arenouGaiaDataRelease2018}; (2) \cite{schutteDiscoveryNearbyYoung2020}; (3) \cite{scholzWhimsAccretingYoung2005}; (4) \cite{cooperUltracoolSpectroscopicOutliers2024}; (5) \cite{boucherBANYANVIIINew2016}; (6) \cite{morales-gutierrezModellingSEDInner2021}.
\label{tab:fundamental_parameters_reshaped}
\end{table*}

\begin{figure}
    \centering
    \includegraphics[width=\linewidth]{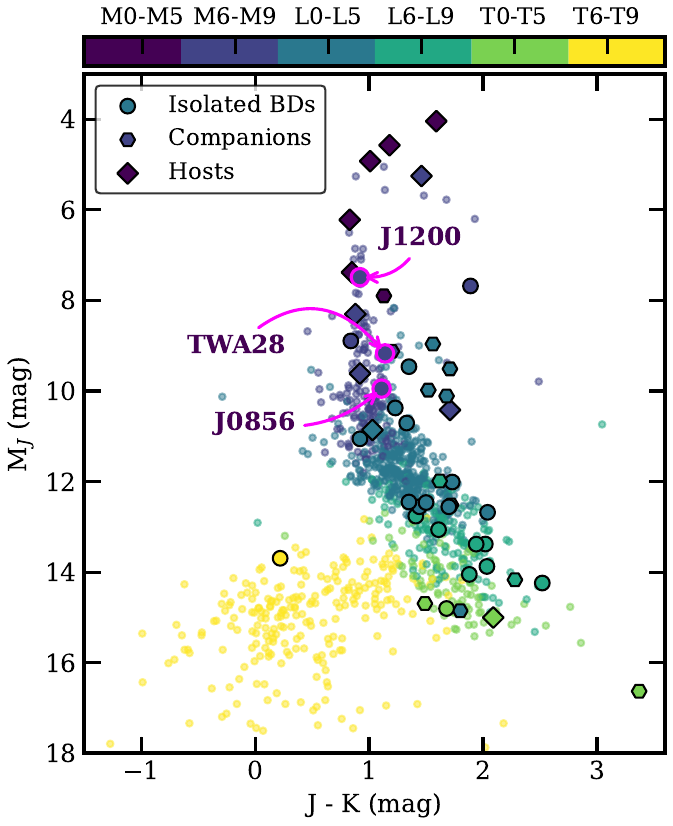}
    \caption{Colour-magnitude diagram of $M_J$ vs $J-K$. The objects observed in the SupJup Survey are indicated with three types of markers corresponding to isolated BDs, substellar companions, and hosts, which are either primary stars or BDs. The population of isolated cool dwarfs as catalogued in the \texttt{UltracoolSheet}\protect\footnotemark is displayed as a reference. The colour of the points indicates the SpTs.}
    \label{fig:color_mag}
\end{figure}

Our sample includes three young ($\lesssim$10 Myr) isolated BDs, with effective temperatures between 2300 and 2800 K and spectral types (SpTs) corresponding to late M-type objects (see Table \ref{tab:fundamental_parameters_reshaped}). Due to ongoing gravitational contraction, these young BDs exhibit low surface gravity ($\log(g) < 4.5$) and low rotational velocities, typically \vsini$ < 15$ \kms \citep{stahlerBirthlineLowmassStars1983,baraffeEvolutionaryModelsLowmass2002,cruzYoungDwarfsIdentified2009,bouvierAngularMomentumEvolution2014}. The selection of these three objects is based on their shared properties with young SJs, including temperature, mass, and age (see Fig. \ref{fig:color_mag}). The detection of SJs with direct-imaging techniques is sensitive to self-luminous objects, making young and massive companions prime targets for detection and characterisation (e.g. \citealt{mcelwainFirstHighContrastScience2007,viganExoplanetCharacterizationLong2008}).

\subsection{J1200}
2MASS J12003792-7845082, hereafter referred to as J1200, is a late M-type BD with a circumstellar disk, as detailed by \citep{schutteDiscoveryNearbyYoung2020}. Located within the $\varepsilon$ Cha association, J1200 is a member of this young ($\sim 3.7$ Myr) stellar group. In their comprehensive analysis involving spectral energy distribution (SED) fitting and near-infrared spectroscopy, \citet{schutteDiscoveryNearbyYoung2020} establishes its SpT as M6$\gamma$ with an effective temperature of approximately 2800 K and a mass ranging between 42 and 58 \Mjup. The assigned spectral subtype $\gamma$ is indicative of low surface gravity, a commonly observed characteristic in young BDs.

\subsection{TWA 28}
2MASS J11020983-3430355, also known as TWA 28, is an accreting substellar member of the dynamic TW Hydrae association (TWA), a young stellar group with an age ranging from 5 to 10 Myr (see \citealt{scholzWhimsAccretingYoung2005} and \cite{mamajekMovingClusterDistance2005}). Analysis of medium-resolution near-infrared spectroscopy conducted by \citet{venutiXshooterSpectroscopyYoung2019} point to an M9-type, corresponding to an effective temperature of \Teff=$2600 \pm 70$ K. The measured surface gravity \logg= 4.1$\pm$0.3 is consistent with low-surface gravity, while the estimated rotational velocity is \vsini = $25\pm10$ \kms. Additionally, the mass and radius are estimated via the accretion luminosity to 20.9$\pm5$ \Mjup{} and $2.8$ \Rjup, respectively. Low-resolution spectroscopy from \textit{Gaia} Data Release 3 \citep{arenouGaiaDataRelease2018} enabled the analysis of a large sample of ultracool dwarfs as reported in \cite{cooperUltracoolSpectroscopicOutliers2024}, where TWA 28 is identified as a low-gravity object with a SpT of M8.5$\gamma$ and an effective temperature of 2382 $\pm$ 42 K, suggesting a lower temperature than previously measured.

Recent observations with the NIRSpec instrument aboard JWST, as presented in \cite{manjavacasMediumresolution97Mm2024}, compared the 0.97-5.3 \micron{} spectra of TWA 28 to models of substellar atmospheres to constrain its atmospheric properties. The authors found that TWA 28 is best fit by models with effective temperatures of 2400-2600 K and a surface gravity of $\log(g) = 4.0 \pm 0.5$. Additionally, the authors report the detection of excess flux in the $>2.5$ micron region, suggesting the presence of a disk around TWA 28, compatible with the accretion signatures found by \cite{venutiXshooterSpectroscopyYoung2019}

\subsection{J0856}

2MASS J08561384-1342242, referred to as J0856, is a young ($10\pm3$ Myr) M8$\gamma$ BD with an estimated mass of $14.4 \pm 1.4$ \Mjup{} \citep{boucherBANYANVIIINew2016}. J0856 shows distinct low-gravity signatures and infrared excess flux, indicating the presence of a disk \citep{boucherBANYANVIIINew2016}. The properties of J0856 closely resemble those of TWA 28, hence providing a suitable comparison object.

Analysis of the SED by \cite{morales-gutierrezModellingSEDInner2021} supports the presence of extended infrared excess emission. The authors modelled its SED with a central source of $T_{\text{eff}} \approx$ 2500 K and a disk. The contribution from the disk becomes comparable to that of the BD photosphere at 4-5 \micron{} and dominant at longer wavelengths.

In the study by \cite{cooperUltracoolSpectroscopicOutliers2024}, the parameters of J0856 parameters were refined to a SpT of M8.6$\beta$ and an effective temperature of 2380 $\pm$ 32 K. The $\beta$ spectral index indicates intermediate surface gravity, consistent with the object's young age but not as low as the $\gamma$ index.

\begin{table*}
    \centering
    \caption{Observational properties of the young BDs.}
    \renewcommand{\arraystretch}{1.5}
    \begin{tabular}{llllllll}
\hline
 System Name             & RA          & Dec.          & K (mag)        & Integration Time (s) & S/N & Date of Observation \\
\hline
 2MASS J12003792-7845082 & 12 00 37.94 & -78 45 08.28 & 11.6 & 3000       & 18.5 & 2023-03-04 \\
 TWA 28                  & 11 02 09.84 & -34 30 35.56 & 11.9 & 4800      & 20.6 & 2023-03-03 \\
 2MASS J08561384-1342242 & 08 56 13.84 & -13 42 24.27 & 12.5 & 4800       & 16.3 & 2023-03-04 \\
\hline
\end{tabular}
    \tablefoot{The signal-to-noise ratio (S/N) is calculated at 2346 nm. The apparent \textit{K}-band magnitudes are from the 2MASS catalogue and have an associated uncertainty of 0.02 mag \citep{cutriVizieROnlineData2003}.}
    \label{tab:observations}
\end{table*}

\section{Observations and data reduction}\label{sec:observations}
\subsection{Observations}
The observations presented in this work were conducted as part of the SupJup Survey (PI: Snellen, \citealt{deregtESOSupJupSurvey2024a}), employing the CRIRES$^+$ instrument at the Very Large Telescope (VLT). CRIRES$^+$, a state-of-the-art high-resolution slit-spectrograph equipped with adaptive optics, covers a broad wavelength range of the infrared spectrum (0.95-5.3 micron), at a resolving power $R \sim 100,000$ \citep{dornCRIRESSkyESO2023}. The data were collected March 3-4, 2023. Of the two observing nights, the first one experienced moderate seeing conditions (1.0-1.5") and high humidity, while on the second night the seeing was stable ($<1.0$").

For this work, we opted for the K2166 wavelength setting, covering from 1.90 to 2.48 $\mu$m, specifically targeting regions rich in \twelveCO{} and, more specifically, \thirteenCO{} absorption features. The observations employed the wide slit mode of 0.4", achieving a resolving power of $R \approx 50,000$ as measured by the line shape of the telluric lines (see \Cref{app:molecfit}). Each target's total exposure time was calibrated to reach a S/N per pixel greater than $15$ in the continuum (see \Cref{tab:observations}). Observations were conducted in nodding mode, with individual exposures lasting 300s (total exposure time for each object outlined in \Cref{tab:observations}), as this allows the correct subtraction of the sky thermal emission. The observations of telluric standard stars were interleaved with the science observations, with the goal of obtaining representative spectra of the Earth's transmission. The telluric standards were observed at a similar airmass and with the same instrumental setup as the science targets. 
\subsection{Data reduction}\label{sec:data_reduction}
The data reduction was performed with the open-source package \textit{excalibuhr}\footnote{\url{https://github.com/yapenzhang/excalibuhr}} (Zhang et al., in prep, also described in \cite{deregtESOSupJupSurvey2024a}). The reduction started by tracing spectral orders on the flat-field images and determining the slit curvature via Fabry-Perot etalon frames, which provides an initial wavelength calibration. Subsequent steps included applying flat-fielding corrections, replacing bad pixels, fixing the slit curvature and tilt, and removing the sky emission through AB (or BA) subtraction. Frames from identical nodding positions were co-added into master frames for both A and B positions, which were then combined into a singular master frame. 
The conversion of calibrated images to one-dimensional spectra utilised the optimal extraction algorithm \citep{horneOPTIMALEXTRACTIONALGORITHM1986}. This procedure was similarly applied to standard star observations. The refinement of wavelength solutions for the extracted spectra was achieved through chi-square minimisation with telluric transmission templates provided by ESO's \textit{Skycalc}\footnote{\url{https://skycalc-ipy.readthedocs.io}} \citep{leschinskiSkyCalc_ipySkyCalcWrapper2021}.

The imprint of the Earth's atmosphere on the observed spectra was corrected for with \texttt{Molecfit} \citep{smetteMolecfitGeneralTool2015} and observations of standard stars. Regions with saturated or deep telluric features were masked and ignored in the subsequent analysis (see \Cref{app:molecfit} for more details). The telluric fit to the standard star also provides the instrumental throughput via a third-degree polynomial fit, which is used to correct the observed spectra for instrumental effects. The telluric-corrected spectra, containing 21 datasets (7 spectral orders with 3 detectors each), are divided by the mean flux of each order-detector pair, resulting in a normalised spectrum.

\footnotetext{\url{http://bit.ly/UltracoolSheet}}

\section{Forward modelling}\label{sec:forward_modeling}
To generate high-resolution spectra of BD atmospheres, we employed the \texttt{petitRADTRANS} code \citep{mollierePetitRADTRANSPythonRadiative2019}. In this framework, the atmosphere is divided into a series of layers covering the pressure range from $10^2$ to $10^{-5}$ bar, providing comprehensive coverage of the atmospheric vertical extent, including the region where spectral lines originate: the photosphere. Each layer is parameterised by a temperature and mass fractions of pertinent chemical species. At each layer, the radiative transfer equation is solved to obtain the outgoing flux as a function of wavelength. The resulting spectra are convolved with rotational and instrumental broadening (see \Cref{app:molecfit}) kernels and resampled to the observed wavelength grid for direct comparison with the data.

Primary input for \texttt{petitRADTRANS} is the pressure-temperature profile, the opacities of spectrally active species, mass fractions of relevant chemical species and the surface gravity of the object. Additionally, the object's radius can be utilised to scale the model to the observed flux. 

\subsection{Pressure-temperature profile}
Self-luminous objects, devoid of significant external irradiation, exhibit a thermal profile expected to decrease with altitude. We parameterised the temperature as a function of pressure using values for the temperature gradient $\nabla T_i \equiv \frac{d\ln T_i}{d\ln P_i}$, where $i=1,2,3,4,5$ corresponds to equally spaced points in pressure log-space. The temperature at each atmospheric layer was calculated as described in \cite{zhangELementalAbundancesPlanets2023a}:

\begin{align*}
T_1 &= T_{\text{bottom}} \quad \text{at} \quad P_1 = 10^2 \quad\text{bar}\\
T_{j+1} &= \exp\left(\ln T_j + (\ln P_{j+1} - \ln P_j) \cdot \nabla T_j \right).
\end{align*}

\begin{figure}
    \centering
    \includegraphics[width=\linewidth]{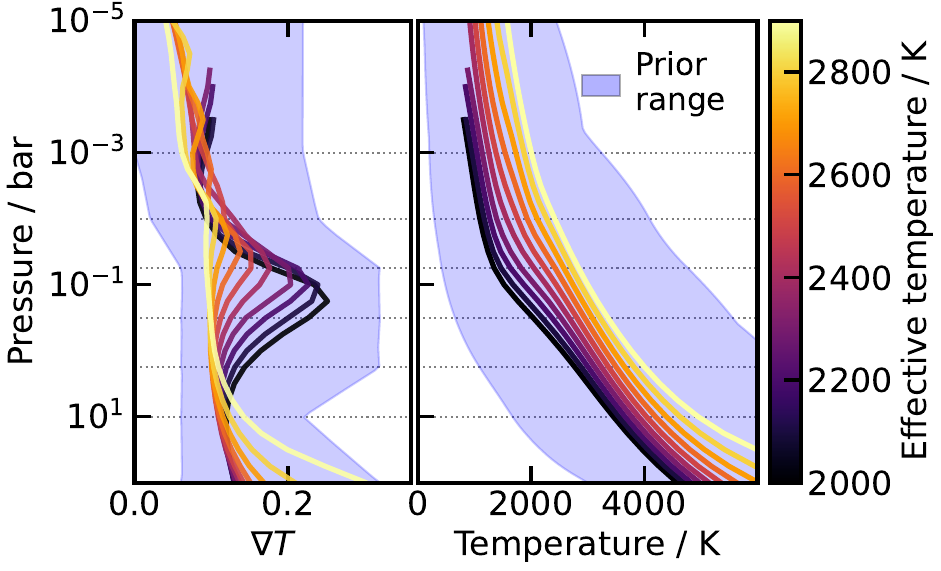}
    \caption{Thermal profiles (right) and temperature gradients (left) of the \texttt{SPHINX} model grid (\url{https://zenodo.org/records/7416042}) covering the range of temperatures between 2000 and 3000 K. The blue envelopes show the parameter space covered by the selected priors for the temperature gradients.}
    \label{fig:sphinx_PT_profiles}
\end{figure}

\noindent Here, $j=1,...,100$ represents the atmospheric layers used in the radiative transfer calculation. The value of $\nabla T_j$ at each layer is determined with linear interpolation from the values at the knots $\nabla T_i$. This approach facilitates the coupling of self-consistent radiative-convective-equilibrium (RCE) models with a `free' parameterisation through custom prior probabilities. The uniform priors for $\nabla T_i$ are set to replicate RCE thermal profiles (see \Cref{fig:sphinx_PT_profiles}). For this study, we considered the temperature profiles derived from state-of-the-art atmospheric models for M dwarfs (\texttt{SPHINX}; \cite{iyerSPHINXMdwarfSpectral2023}). The \texttt{SPHINX} profiles closely match those of other theoretical models in overlapping temperature ranges; however, the lack of high-temperature models of cool, substellar objects \citep{marleySonoraBrownDwarf2021,mukherjeeSonoraSubstellarAtmosphere2024} makes \texttt{SPHINX} the most suitable framework to derive physical constraints on the temperature profiles of the objects of interest in our study: late-M BDs. The prior probability distribution for each gradient $\nabla T_i$ and for the temperature at the bottom of the atmosphere is determined such that the resulting temperature profiles are consistent with the \texttt{SPHINX} models. In this way, the priors of the parameters are informed by RCE profiles, hence coupling the flexibility of the retrieval framework with the physical constraints of the models. Our parameterisation is expanded by including a free parameter that shifts the position of the intermediate pressure-temperature knots in the log-pressure space, denoted as $\log \Delta P$. This approach allows us to explore temperature profiles without biasing the photosphere's location, which is expected to vary with the object's surface gravity and effective temperature.

\subsection{Chemical composition}
\begin{figure}
    \centering
    \includegraphics[width=\linewidth]{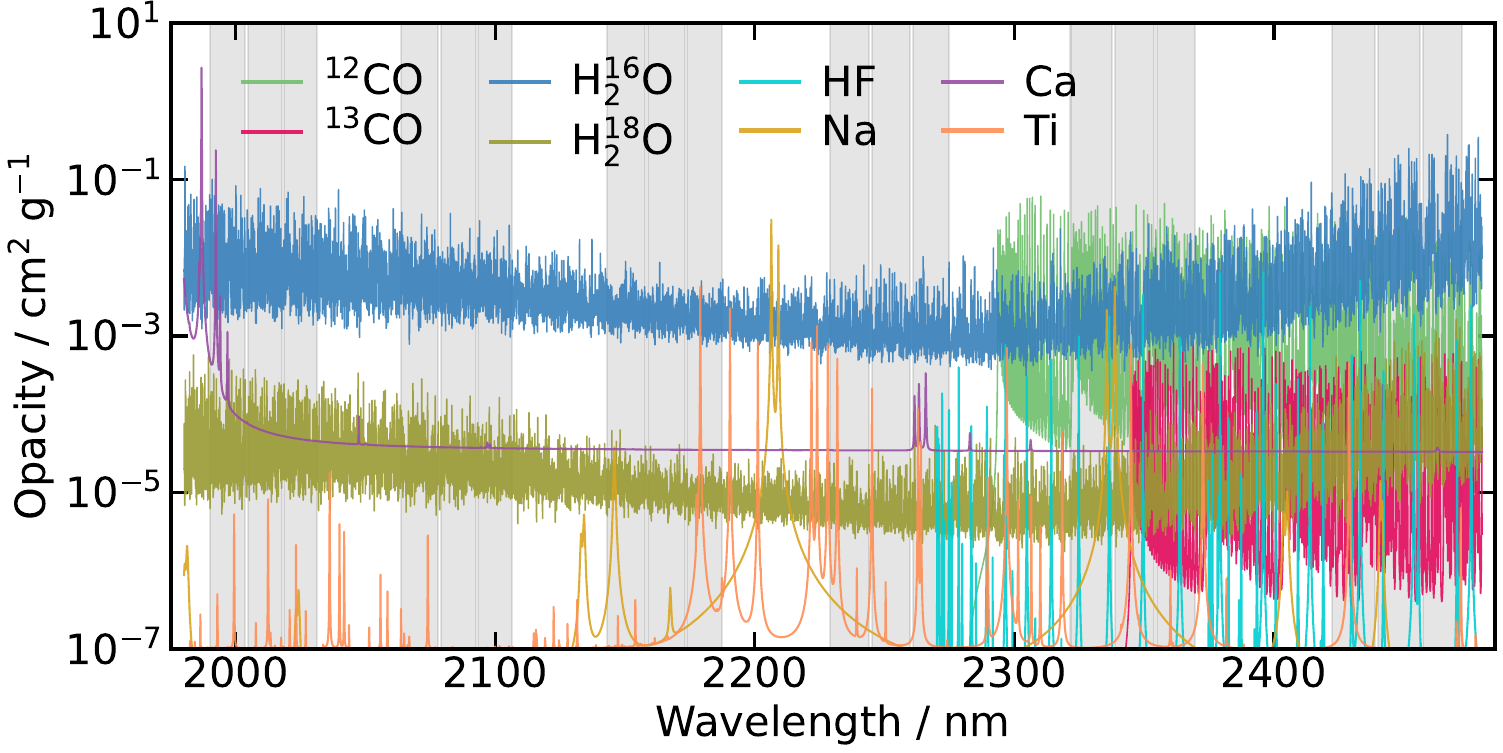}
    \caption{Opacity sources included in the forward modelling. The opacities are shown for a temperature of 2400 K and at the retrieved mixing ratios of TWA 28 (see \Cref{tab:retrieval_results}).}
    \label{fig:opacity_sources}
\end{figure}

The temperature range of 2000-3000 K straddles the boundary between planetary photospheres and those of low-mass stars. The spectra of objects within this temperature range are predominantly characterised by molecular lines (see Fig. \ref{fig:opacity_sources}), with some atomic lines as seen in stars from species such as Na, K, Ca, Fe, and Ti \citep{cushingInfraredSpectroscopicSequence2005}.

In the K band ($\lambda$1.9-2.5 $\mu$m), the main spectral features arise from rotational-vibrational transitions of water and carbon monoxide. Dense water lines are pervasive across the observed wavelength range, while CO features are concentrated in the two reddest spectral orders of CRIRES$^+$ ($\lambda$2.29-2.48 \micron). Notable CO isotopologue overtones covered in our observations include \thirteenCO$_{(2,0)}$ ($\lambda2.345$ \micron) and \thirteenCO$_{(4,2)}$ ($\lambda2.403$ \micron) \citep{harrisonDeterminations12132017}.

Opacities used in this work were calculated utilising the latest line lists from ExoMol \citep{tennysonExoMolDatabaseMolecular2016} for \water{} \citep{polyanskyExoMolMolecularLine2017,polyanskyExoMolMolecularLine2018}, HITEMP for the CO isotopologues \citep{rothmanHITEMPHightemperatureMolecular2010, liROVIBRATIONALLINELISTS2015}, and for atomic opacities: Na, K \citep{allardNewStudyLine2019}, Mg, Ca, Ti, and Fe \citep{castelliNewGridsATLAS92004}. Additionally, we included hydrogen fluoride (HF) as a line species \citep{wilzewskiH2HeCO22016}. Relevant continuum opacities encompass \Htwo-\Htwo, \Htwo-He, H- \citep{dalgarnoRayleighScatteringMolecular1962,chanRefractiveIndexHelium1965, grayObservationAnalysisStellar2022}, and collision-induced absorption \citep{borysowCollisioninducedRototranslationalAbsorption1988}.

Our retrieval framework treats the chemical abundance of each opacity source as an independent, free parameter. In contrast to equilibrium chemistry (EC), this approach avoids imposing physical constraints on the relative abundances among different species. This method enhances sensitivity to individual species and is less influenced by assumptions inherent in atmospheric models.

The number fractions, $n_X$, of the main carbon- and oxygen-bearing species, CO and H2O respectively, are used to derive the atmospheric C/O for each object (see Fig. \ref{fig:bestfit_chemistry}), defined as
\begin{align}\label{eq:carbon_to_oxygen}
    \mathrm{C/O} = \frac{n_{\mathrm{CO}}}{n_{\mathrm{H_2O}}+n_{\mathrm{CO}}}= \frac{n_{\rm \twelveCO} + n_{\rm \thirteenCO}}{n_{\rm \sixteenOwater} + n_{\rm \eighteenOwater} + n_{\rm \twelveCO} + n_{\rm \thirteenCO}}.
\end{align}
We note that the C/O might change as a function of altitude if there are vertical gradients in the abundances of the species.\ In this work we assumed constant-with-altitude abundances and, hence, a constant C/O. The atmospheric C/O might differ from the bulk C/O of the object, as cloud formation can result in the depletion of oxygen in the deep atmosphere \citep{lineUniformAtmosphericRetrieval2015}. At the hot temperatures of our objects, most condensates are expected to be in the gas phase; however, there are a few species, such as Al$_2$O$_3$ and Mg$_2$SiO$_4$, that can form condensates at temperatures above 2000 K \citep{hellingAtmospheresBrownDwarfs2014,woitkeDustBrownDwarfs2020}. The presence of clouds can affect the slope of the continuum and the shape of the temperature profile. Our \textit{K}-band observations cover a small wavelength region and are not sensitive to the overall shape of the continuum (see Sect. \ref{sec:data_reduction}); hence, we do not expect the presence of clouds to significantly affect our results.

As a proxy for metallicity, we employed the carbon abundance relative to hydrogen scaled to solar composition as
\begin{align}
    \mathrm{[C/H]} = \log_{10}\left(\frac{n_{\mathrm{C}}}{n_{\mathrm{H}}}\right) - \log_{10}\left(\frac{n_{\mathrm{C}}}{n_{\mathrm{H}}}\right)_{\odot},
\end{align}
where the solar value is $\log_{10}\left(\frac{n_{\mathrm{C}}}{n_{\mathrm{H}}}\right)_{\odot} = -3.57$ \citep{asplundChemicalMakeupSun2021}. In a similar manner, the fluorine abundance is calculated from HF as
\begin{align}
    \mathrm{[F/H]} = \log_{10}\left(\frac{n_{\mathrm{HF}}}{n_{\mathrm{H}}}\right) - \log_{10}\left(\frac{n_{\mathrm{HF}}}{n_{\mathrm{H}}}\right)_{\odot},
\end{align}
where the solar value is $\log_{10}\left(\frac{n_{\mathrm{F}}}{n_{\mathrm{H}}}\right)_{\odot} = -7.6 \pm 0.25$ \citep{maiorcaNEWSOLARFLUORINE2014, asplundChemicalMakeupSun2021}.

To ascertain carbon isotope ratios, we retrieved the number fractions of \twelveCO{} and \thirteenCO{} independently. The \Cratio\ is then calculated as
\begin{align}
    \mathrm{\Cratio} = \frac{n_{\mathrm{\twelveCO}}}{n_{\mathrm{\thirteenCO}}},
\end{align}
similarly, the \ratio{O}{16}{18} ratios are calculated as
\begin{align}
    \mathrm{\ratio{O}{16}{18}} = \frac{n_{\mathrm{\sixteenOwater}}}{n_{\mathrm{\eighteenOwater}}}.
\end{align}

To robustly confirm the presence of molecular species, we conducted a cross-correlation check following the methodology outlined in \citep{zhang13COrichAtmosphereYoung2021}. The cross-correlation function (CCF) for each species $X$ was calculated as 

\begin{align}
    \mathrm{CCF}_X(v) = \sum_{i=1}^{N_{\lambda}} \frac{(d_i - m_{i, \tilde{X}})(m_i - m_{i, \tilde{X}})}{\sigma_i^2},
\end{align}

\noindent where $m_{i, \tilde{X}}$ is the model without the species $X$, and $m_i$ is the model containing all species included in the retrieval. The CCF is calculated for shifted versions of the model spectrum in velocity space covering $v=(-1000, 1000)$ \kms{} with steps of 1 \kms{}. The uncertainties $\sigma_i$ are calculated from the diagonal values of the best-fit covariance matrix, which accounts for the uncertainty of each data point and correlated noise among nearby pixels (see Sect. \ref{subsec:correlated_noise}).

\subsection{Line profile}
Several physical mechanisms contribute to the observed line spread function. Intrinsic rotation of the object induces Doppler-shifted lines across different regions of the observed disk, leading to an overall spectrum with broadened spectral lines. To account for this, we utilised the \texttt{fastRotBroad} broadening kernel from \texttt{PyAstronomy}\footnote{\url{https://github.com/sczesla/PyAstronomy}} \citep{czeslaPyAPythonAstronomyrelated2019}. This kernel is employed to broaden our model lines based on the projected rotational velocity \vsini, where $i$ denotes the inclination of the system, and $\varepsilon_{\text{limb}}$ represents the linear limb-darkening coefficient. It is important to note that \texttt{fastRotBroad} assumes a constant-with-wavelength broadening kernel for computational efficiency, a valid approximation within our small wavelength range ($\Delta \lambda \approx 50$ nm), showing a negligible difference (<0.5\%) with direct disk integration \citep{carvalhoSimpleCodeRotational2023}.

To correct for the shift in line positions induced by the radial velocity ($v_{\text{rad}}$) of the object, we shift the forward-modelled spectra via linear interpolation to match the observed Doppler shift. Furthermore, to incorporate the instrumental broadening introduced by the slit spectrograph, we convolve the lines with a Gaussian kernel. The width of this kernel is determined by the full width  at half maximum (FWHM), which corresponds to the spectral resolution of the instrument ($R \sim 50,000$; see \Cref{app:molecfit}).

\subsection{Veiling}\label{subsec:veiling_continuum}
Classical T Tauri stars often exhibit veiling of photospheric lines, characterised by shallower absorption features than expected for their SpT \citep{hartiganHowUnveilTauri1989, stempelsPhotosphereVeilingSpectrum2003,sullivanOpticalNearinfraredExcesses2022,sousaNewInsightsNearinfrared2023}. Veiling is attributed to the presence of a circumstellar disk, which can introduce an additional continuum source, frequently observed as an infrared excess \citep{mcclureCHARACTERIZINGSTELLARPHOTOSPHERES2013}.

In our analysis, veiling is modelled as an additional continuum component to the observed flux, with wavelength slope and amplitude as free parameters. This component is included in the forward model and retrieved simultaneously with other parameters. The veiling continuum is expected to be stronger for objects with accretion signatures. We do not observe any emission lines in our sample that would be indicative of active accretion, such as Br$\gamma$ emission \citep{beckSpatiallyExtendedBrackett2010}.

The veiling continuum is modelled as a power-law function of the form

\begin{align}
    \mathbf{d}(\lambda) &= \phi[\mathbf{m}_{\text{spec}}(\lambda) + \mathbf{m}_{\text{veil}}(\lambda)] \\&= \phi \left[\mathbf{m}_{\text{spec}}(\lambda) + r_0 \left(\frac{\lambda}{\lambda_0}\right)^{\alpha}\right],
\end{align}
where $\phi$ is a normalising constant fitted for each order-detector pair (see Sect. \ref{subsec:likelihood}), $r_0$ is the veiling factor at the shortest wavelength of our dataset ($\lambda = 1.90$ \micron), and $\alpha$ is the power-law index.

In this parameterisation, the wavelength-dependent \textit{K}-band veiling factor, $r_k$, is
\begin{align}
\mathbf{r_k}(\lambda) = r_0 \left(\frac{\lambda}{\lambda_0}\right)^{\alpha},
\end{align}
which is a quantity often reported in the literature, with values ranging from 0 (no veiling) to of the order of 5 for very active T Tauri stars \citep{reiLinedependentVeilingVery2018,alcalaGIARPSHighresolutionObservations2021,sousaNewInsightsNearinfrared2023}.

\begin{figure*}[h!]
    \centering
    \includegraphics[width=\linewidth]{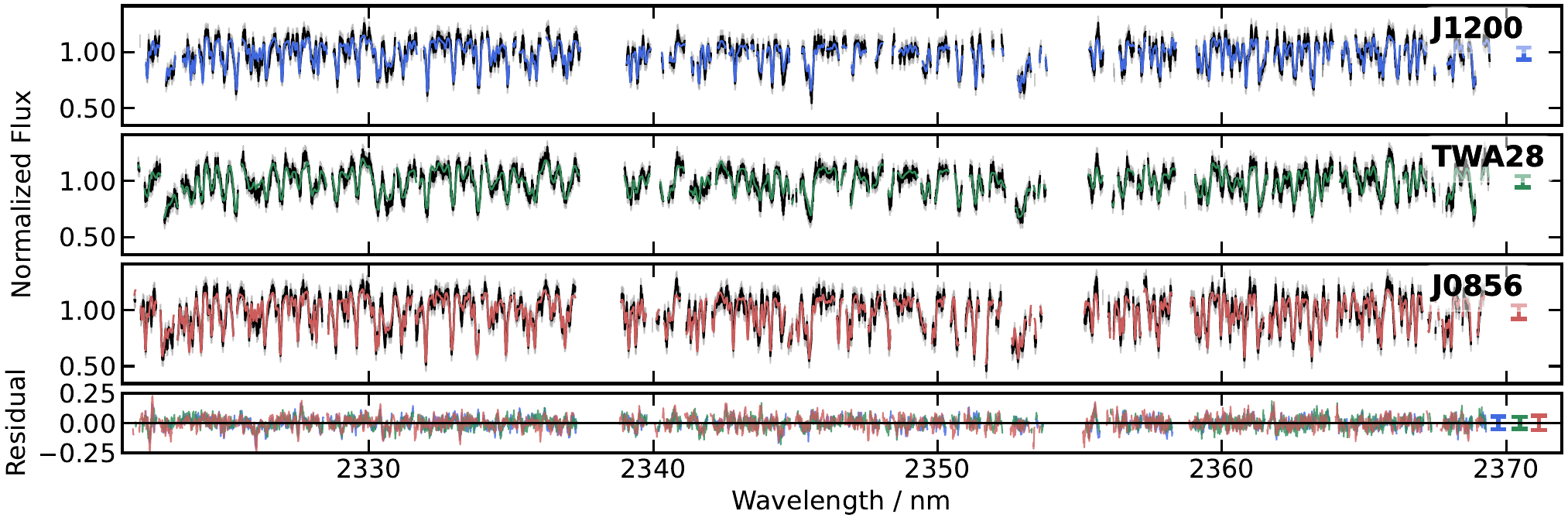}
    \caption{Best-fitting models retrieved for the three objects. The spectra correspond to the three detectors of a single order of \CRIRES. This wavelength range contains several \twelveCO{} and \thirteenCO{} lines. The observed data are shown in black, and the corresponding best-fit models are in blue (J1200), green (TWA 28), and red (J0856). Residuals are shown in the bottom panel. The average scaled uncertainty is indicated on the right side of each panel. The data and the models are displayed in the rest frame of each object. The full wavelength coverage used in the retrievals is shown in \Cref{app:bestfit_spectra}.}
    \label{fig:bestfit_spectra}
\end{figure*}

\begin{figure*}
    \centering

    \begin{subfigure}{0.45\textwidth}
        \includegraphics[width=\linewidth]{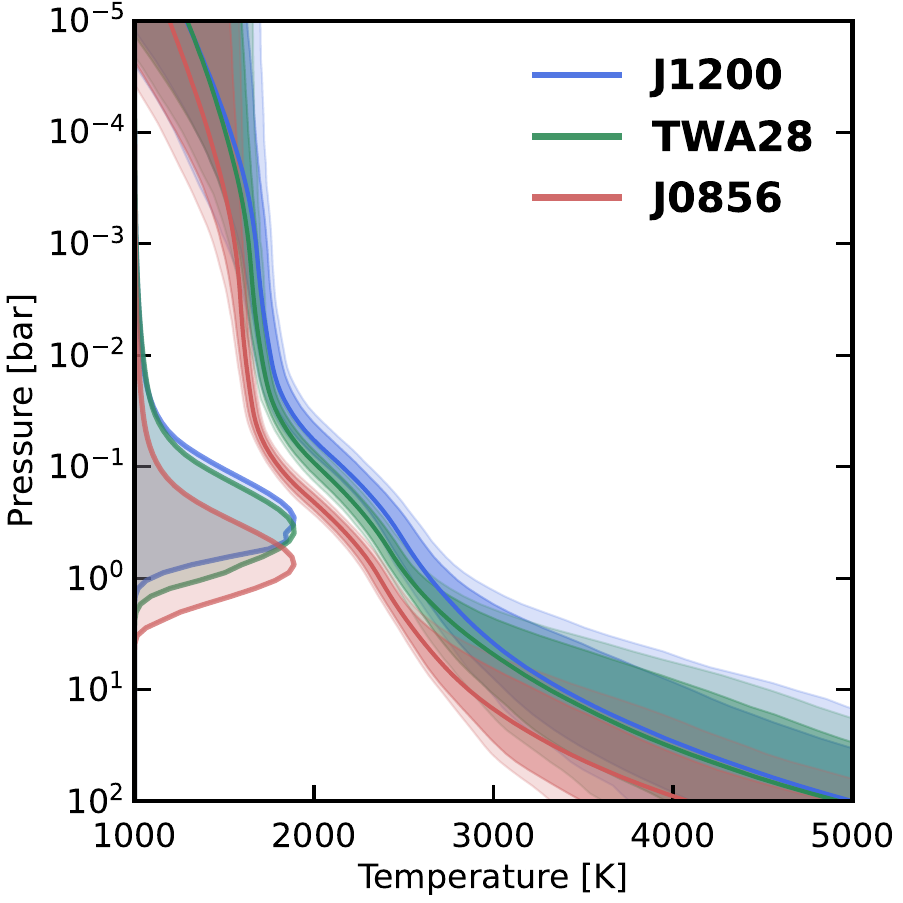}
        \caption{}
        \label{fig:bestfit_PT}
    \end{subfigure}
    \begin{subfigure}{0.45\textwidth}
        \includegraphics[width=\linewidth]{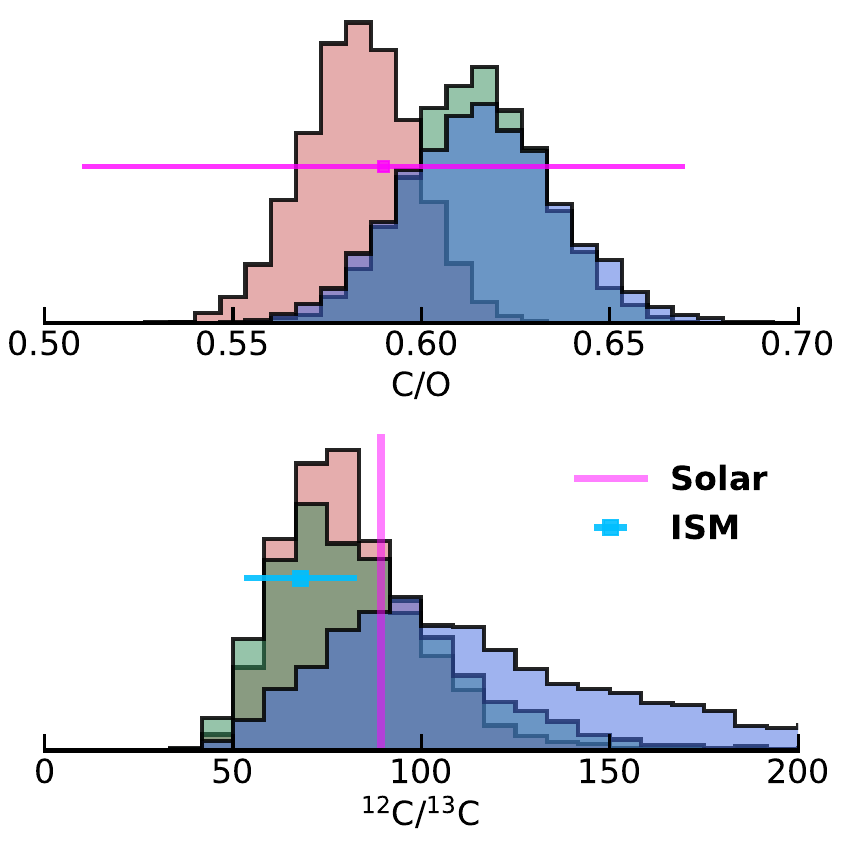}
        \caption{}
        \label{fig:bestfit_chemistry}
    \end{subfigure}
    \caption{(a) Best-fit pressure-temperature profiles. The shaded regions indicate the 1-, 2-, and  3-sigma regions. On the right axis we overplot the integrated contribution function. (b) Posterior distributions for C/O (top) and \Cratio\ (bottom). The values for the solar C/O $=0.59\pm0.08$ \citep{asplundChemicalMakeupSun2021}, and the isotope ratios for the Solar System, $89.3\pm0.2$ \citep{meijaIsotopicCompositionsElements2016}, and the ISM, \Cratioism = $68 \pm 16$ \citep{milam1213Isotope2005}, are marked.}
\end{figure*}

\section{Atmospheric retrieval framework}\label{sec:retrieval_framework}

Atmospheric retrievals aim to solve the inverse problem of finding the best-fitting parameters and their uncertainties.
We employed a Bayesian inference framework for our retrieval process, allowing us to quantify the uncertainties in the derived atmospheric parameters. We used the nested sampling algorithm (\citealt{skillingNestedSampling2004}), implemented through \texttt{MultiNest} \citep{ferozMultiNestEfficientRobust2009} with the Python wrapper \texttt{PyMultiNest} \citep{buchnerPyMultiNestPythonInterface2016}. \texttt{MultiNest} is adept at efficiently exploring and sampling from complex, high-dimensional parameter spaces, effectively managing parameter degeneracies. The sampling employs a set of live points drawn from the hyper-dimensional prior parameter space, defining an iso-likelihood contour. Through repeated likelihood evaluations (refer to Sect. \ref{subsec:likelihood}), a fraction of live points is iteratively replaced to shrink the parameter space to regions of higher likelihood. We made use of the importance nested sampling technique in constant efficiency mode, which provides reasonably accurate evidence (\evidence) estimates at significantly higher efficiency than traditional nested sampling \citep{cameronRecursivePathwaysMarginal2014, ferozImportanceNestedSampling2019}. Our retrievals operate with 400 live points at a constant efficiency of 5\%, with a tolerance of $\Delta\log$\evidence=0.5, following the recommendations in \citep{ferozImportanceNestedSampling2019}. Additional retrievals with different numbers of live points and efficiencies were performed to ensure convergence and consistency of the results. We note that the results presented here are consistent with the number of live points varying from 200 to 1000 and an efficiency of 1-5\% (see \cite{dittmannNotesPracticalApplication2024} for a detailed discussion on \texttt{MultiNest}'s hyperparameters).

\subsection{Likelihood function}\label{subsec:likelihood}
The likelihood function, as defined in \cite{ruffioRadialVelocityMeasurements2019} and employed in \cite{wangRetrievingAbundancesHR2022,landmanPictorisEyesUpgraded2024}, is expressed for a data vector $d_{ik}$ of shape ($N_i$, $N_k$) = (21, 2048), corresponding to the combination of 7 spectral orders and 3 detectors (totalling 21 order-detector pairs) with 2048 pixels each. A model $M_{\psi, ik}$ with non-linear parameters $\psi$ and amplitudes $\phi_{i}$ is used. The log-likelihood is computed as follows:
\begin{equation}
    \ln \mathcal{L} = \sum_{i} \ln \mathcal{L}_{i}
\end{equation}
\begin{equation}
    \ln \mathcal{L}_{i} = -\frac{1}{2}\left(N_k \ln(2\pi) + \ln(|\Sigma_{0,i}|) + N_k \ln(s_{i}^2) + \frac{1}{s_{i}^2} \mathbf{r}_{i}^T \Sigma_{0,i}^{-1} \mathbf{r}_{i}\right)
,\end{equation}
where $\Sigma_0$ is the covariance matrix, $\mathbf{r}_{i} = d_{ik} - \phi_{i}m_{ik}$ represents the residuals for each order-detector pair, and $s_{i}$ is the uncertainty scaling factor for the total covariance matrix $\Sigma_{i} = s_{i}^2 \Sigma_{0,i}$.

The optimal parameters $\tilde{\phi}_{i}$ and $\tilde{s}_{i}^2$ are are calculated at every likelihood evaluation using a least square solver\footnote{\url{https://docs.scipy.org/doc/scipy/reference/generated/scipy.optimize.nnls.html}} and the uncertainty scaling formula, respectively:
\begin{equation}
    \tilde{\phi}_{i} = (M\Sigma_0^{-1}M)^{-1}M\Sigma_0^{-1}\mathbf{d}
\end{equation}
\begin{equation}
    \tilde{s}_{i}^2 = \frac{1}{N_k}\mathbf{r}_{i}^T \Sigma_{0,i}^{-1} \mathbf{r}_{i}\bigg|_{\phi=\tilde{\phi}}
.\end{equation}

\subsection{Correlated noise}\label{subsec:correlated_noise}
Uncertainties in the data play a crucial role in the retrieval process. Recent findings by \cite{deregtESOSupJupSurvey2024a} underscore the significance of correlated noise in CRIRES$^+$ spectra on the retrieved parameters. When uncertainties are correlated among nearby pixels, the covariance matrix $\Sigma_0$ is not diagonal. Gaussian processes (GPs) provide a powerful means to model such correlated noise \citep{kawaharaAutodifferentiableSpectrumModel2022}. In this study, we employed a radial basis function kernel to model correlated noise in the data, defined as
\begin{equation}
    k_{ij} = a^2 \sigma_{\text{eff}, ij}^2 \exp\left(-\frac{1}{2}\frac{(x_i-x_j)^2}{l^2}\right),
\end{equation}
where $a$ represents the amplitude, $\sigma_{\text{eff}, ij}^2$ the effective variance, calculated as the average variance per order-detector pair, $l$ the length-scale in units of wavelength, and $x_i$  the wavelength of the $i$-th pixel. The resulting covariance matrix for each order-detector pair is the sum of the diagonal covariance matrix and the GP kernel:
\begin{equation}
    \Sigma_{0,ij} = \delta_{ij} \sigma_{i}^2 + k_{ij}.
\end{equation}
During retrieval, the amplitude $a$ and length scale $l$ of the GP kernel are treated as free parameters. The effective variance is derived from the data, and the diagonal covariance matrix corresponds to the uncertainty in the data. The GP kernel is computed at each likelihood evaluation and added to the covariance matrix.

\section{Results}\label{sec:results}

Our best-fitting models successfully reproduce the observed spectral features within the specified uncertainties as shown in Fig. \ref{fig:bestfit_spectra} and quantified via the reduced chi-square (without the noise model, i.e. no uncertainty scaling) in \Cref{app:cornerplot}. The chemical abundances and temperature profiles for the three objects are broadly comparable, with variations attributed to differences in their effective temperatures, surface gravities, spin, and the S/N of our observations. A summary of the retrieved parameters is provided in \Cref{tab:retrieval_results} and a detailed discussion of the results is presented in the following sections.

\subsection{Chemical composition}
Our retrievals provide the best-fit volume-mixing ratios (VMRs) of spectrally active species in the atmospheres of the three objects (see \Cref{tab:retrieval_results}). The derived C/Os according to Eq. \ref{eq:carbon_to_oxygen} are

\begin{align*}
    \mathrm{C/O}_\text{ J1200} & = 0.62^{+0.02}_{-0.02},\\ 
    \mathrm{C/O}_\text{ TWA28} & = 0.61^{+0.02}_{-0.02},\\ 
    \mathrm{C/O}_\text{ J0856} & = 0.58^{+0.01}_{-0.01},\\ 
\end{align*}
where the uncertainties indicate the 1$\sigma$ intervals. The C/Os are comparable among the three targets and close to the solar value of 0.59 $\pm$ 0.08 (\cite{asplundChemicalMakeupSun2021}; see Fig. \ref{fig:bestfit_chemistry}) and similar to the C/O of the young BD 2M0355 ($0.56\pm0.02$; \cite{zhang12CO132021}).

The derived posterior distributions of [C/H] and surface gravity for each object are displayed in Fig. \ref{fig:bestfit_logg_FeH}. Using the carbon-to-hydrogen ratio as a proxy for metallicity, we find that the retrieved metallicities are broadly consistent with the solar value. However, their interpretation is complicated by the high correlation between surface gravities and metallicities (see Sect. \ref{subsec:logg_metallicity}).

In addition to the main molecular species, we detect \thirteenCO, HF, Na, and Ca in the atmospheres of the three objects. In TWA 28 and J0856, we retrieve the VMR of \eighteenOwater{} and tentatively detect it in J1200 (see Sect. \ref{subsec:isotopic_composition}). We also report the presence of Ti in the two hottest targets, J1200 and TWA 28, and provide an upper limit for J0856 (see Fig. \ref{fig:cornerplot_J0856}).

To assess the contribution to the best-fit model and quantify the detection significance of the minor isotopologues, we ran a series of retrievals with the species removed from the model. The significance of the detection is determined via a Bayesian model comparison, where the Bayes factor, $B_{\text{m}}$ \citep{kassBayesFactors1995} is calculated as the ratio of the marginal likelihoods of the models with and without the species
\begin{align}
    \ln B_{\text{m}} = \ln \mathcal{Z}_{\text{m}} - \ln \mathcal{Z}_{\text{m}_{\tilde{X}}}.
\end{align}
The Bayes factor is then converted to a detection significance, with values greater than 3 indicating a significant detection \citep{bennekeHOWDISTINGUISHCLOUDY2013}.

We find that the detection of \thirteenCO{} is strong for TWA 28 and J0856, with 3.9$\sigma$ and 4.7$\sigma$ significance, respectively, while the detection of \eighteenOwater{} is moderate with 2.1$\sigma$ and 3.3$\sigma$ significance for TWA 28 and J0856, respectively. The best-fit model of J1200 exhibits weak evidence for the presence of \thirteenCO{} with a 1.8$\sigma$ significance. The best-fit model for J1200 with \eighteenOwater{} is disfavoured at 1.6$\sigma$; however, the lowest reduced chi-squared ($\chi_{\rm r}$) corresponds to the model with both minor isotopologues (see \Cref{app:cornerplot}). Given the lack of strong evidence of \eighteenOwater{} in J1200, we report the retrieved VMR as an upper limit.

\begin{figure}
    \centering
    \includegraphics[width=\linewidth]{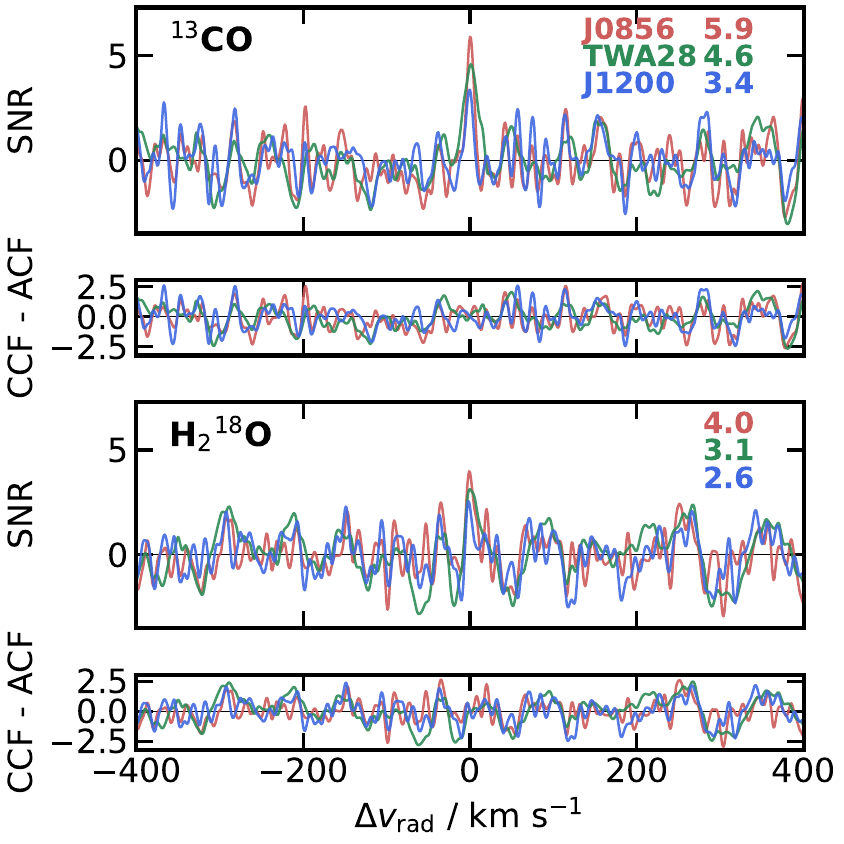}
    \caption{CCFs for \thirteenCO{} and \eighteenOwater{}. The CCFs are calculated for each order-detector pair and summed over all orders and detectors. The CCFs are converted to S/N by dividing them by the standard deviation of the CCFs away from the peak ($|v_{\rm rad}|$ > 100 \kms). The S/N of the peaks of the CCFs is shown in the legend. The residuals between the observed and modelled CCFs  --  the auto-correlation function (ACF) -- are shown in the bottom panel.}
    \label{fig:CCF_13CO_H218O}
\end{figure}

The CCFs support the presence of the minor isotopologues in the atmospheres of the three objects (see Fig. \ref{fig:CCF_13CO_H218O}). The S/Ns of the CCFs are consistent with the Bayesian significances for the strong detections, with a slightly higher detection significance in the cross-correlation analysis. The CCFs of J1200 show a moderate detection of \thirteenCO{} and \eighteenOwater{} at 3.4$\sigma$ and 2.6$\sigma$ level, respectively, higher than the Bayesian significance. We note that in the retrievals without the minor isotopologues, the atmospheric and noise parameters can be adjusted to compensate for the absence of the species; however, the retrieved parameters are consistent to within 1$\sigma$ confidence across all retrievals of the same target. The best fit to the data for all three targets is achieved when the minor isotopologues are included in the model, as reflected in the lowest reduced chi-squared values (see \Cref{app:cornerplot}).

\begin{figure}
    \centering
    \begin{subfigure}{0.9\linewidth}
      \centering
      \includegraphics[width=\linewidth]{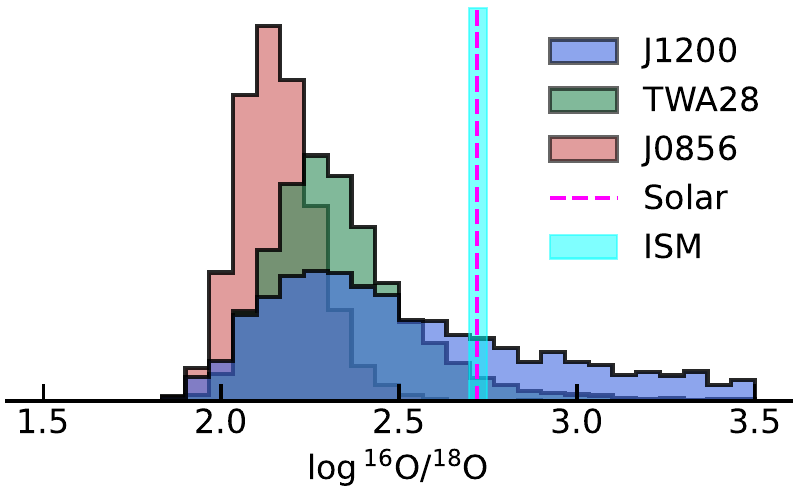}
      \caption{}
      \label{fig:O_ratio}
    \end{subfigure}
    \vskip\baselineskip
    \begin{subfigure}{0.9\linewidth}
      \centering
      \includegraphics[width=\linewidth]{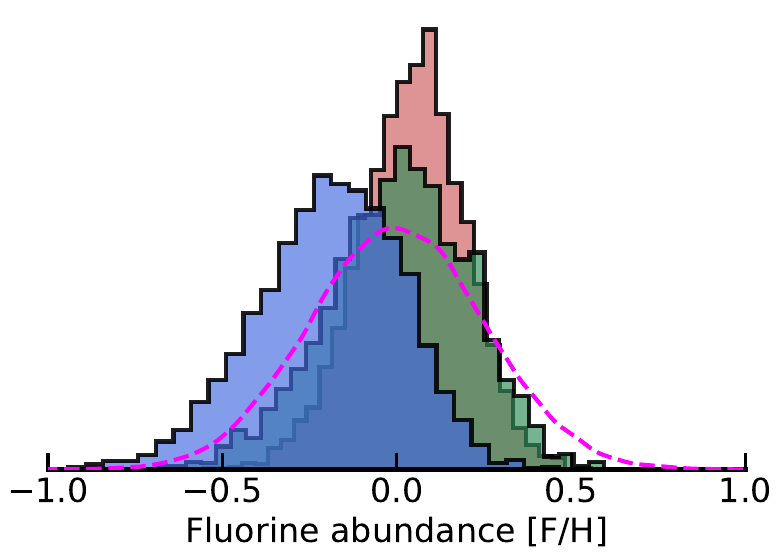}
      \caption{}
      \label{fig:F_H_posterior}
    \end{subfigure}
    \caption{(a) Posterior distributions of the oxygen isotope ratio \ratio{O}{16}{18} for the three objects. The solar value of $525\pm21$ \citep{lyonsLightCarbonIsotope2018} and the ISM value of $557\pm30$ \citep{wilsonIsotopesInterstellarMedium1999} are marked. (b) Posterior distributions of the abundance of fluorine with respect to hydrogen normalised to the solar value. The solar fluorine abundance is overplotted as a Gaussian distribution with a 1$\sigma$ scatter \citep{maiorcaNEWSOLARFLUORINE2014}.}
    \label{fig:all_posteriors}
\end{figure}

\subsection{Thermal profile}
The best-fitting temperature profiles (see Fig. \ref{fig:bestfit_PT}) exhibit remarkable similarities among all three objects, consistent with their similar SpTs and effective temperatures. However, the photospheric region of each object shows slight variations reflecting their individual temperature profiles and surface gravities. The observed spectral lines originate from a region spanning 0.01 to 3 bar, as indicated by the integrated emission contribution function and the narrower uncertainties within this region (see Fig. \ref{fig:bestfit_PT}). 

Moreover, the retrieved thermal profiles align well with literature values for the effective temperature of each object (see \Cref{tab:fundamental_parameters_reshaped}). The inclusion of a free parameter for the spacing of the temperature knots avoids potential biases on the location of the photosphere due to fixed pressure knots and provides a wider range of temperature profiles. The retrieved $\log \Delta P$ values for TWA 28 and J0856 highlight the ability of this parameterisation to capture the location of the photosphere accurately. In contrast, the retrieved $\log \Delta P$ for J1200 is constrained around zero, indicating that even when the fixed pressure knots can already capture the photosphere's location, the additional parameter offers a more accurate representation of the uncertainties in the temperature profile (see \Cref{tab:retrieval_results} and Fig. \ref{fig:bestfit_PT}).

\subsection{Rotational velocity}
For TWA 28, we report a lower value than the previous estimate of $25\pm10$ \kms \citep{venutiXshooterSpectroscopyYoung2019},

\begin{align*}
    v\sin{i}_\text{ TWA28} & = 11.6^{+0.1}_{-0.1}\text{{ km s}}^{{-1}}.
\end{align*}
TWA 28 is the fastest rotator of our sample, in agreement with the observed broadening of the spectral lines (see \Cref{fig:bestfit_spectra,app:bestfit_spectra}). The retrieved \vsini values for J0856 and J1200 are consistent with the expected slow rotation for young BDs \citep{vosSpitzerVariabilityProperties2020}:

\begin{align*}
    v\sin{i}_\text{ J1200} & = 5.3^{+0.2}_{-0.2}\text{{ km s}}^{{-1}},\\ 
    v\sin{i}_\text{ J0856} & = 4.5^{+0.2}_{-0.1}\text{{ km s}}^{{-1}},
\end{align*}

The rotational velocity of low-mass objects is a key parameter in understanding their evolution \citep{bryanWorldsTurnConstraining2020}. Young objects (<1000 Myr) are expected to exhibit slow rotation due to the conservation of angular momentum during the contraction phase \citep{bouvierAngularMomentumEvolution2014}. The observed values for \vsini are are consistent with slow rotation; however, the inclination of the system is unknown, and the true rotational velocity may be higher for a face-on system.

\section{Discussion}
\subsection{Hydrogen fluoride}
Hydrogen fluoride, marked by its 1-0 vibrational transitions with prominent absorption lines in the 2.3-2.5 \micron{} region (see \Cref{app:HF}), is clearly detected in our observations. With precedence in solar observations \citep{hall1969observation} and various giant and dwarf stars \citep{wallace1996high}, the inclusion of HF as a line species stems from its detection in the analysis of high S/N data from a nearby BD (de Regt et al., in prep.). The absolute abundances of fluorine with respect to hydrogen are calculated from the VMRs of HF (see Fig. \ref{fig:F_H_posterior}), 

\begin{align}
    [\rm F/H]_{\text{ J1200}} &= -0.18^{+0.20}_{-0.22},\\
    [\rm F/H]_{\text{ TWA 28}} &= 0.01^{+0.20}_{-0.21},\\
    [\rm F/H]_{\text{ J0856}} &= 0.05^{+0.14}_{-0.15},
\end{align}
which are in excellent agreement with the solar fluorine abundance (see Fig. \ref{fig:F_H_posterior}) as measured from a sunspot spectrum \citep{maiorcaNEWSOLARFLUORINE2014,asplundChemicalMakeupSun2021}.

\subsection{Chemical equilibrium}
\begin{figure}
    \centering
    \includegraphics[width=\linewidth]{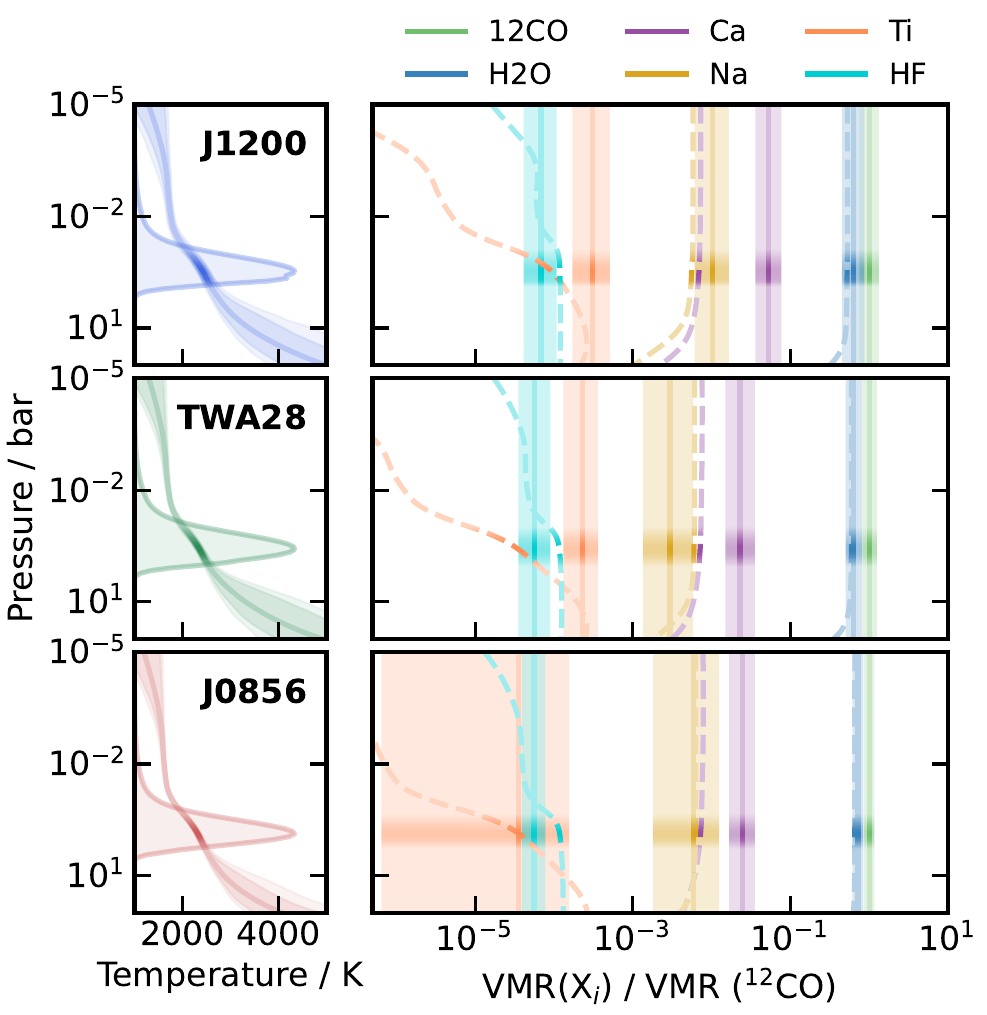}
    \caption{Retrieved VMRs of line species compared to EC values of the retrieved temperature profiles of each target. The dashed lines indicate the values from EC at the retrieved C/O and metallicity. The shaded regions represent the retrieved 1$\sigma$ confidence region around the best fit. The integrated contribution function is plotted in the left column and indicated in the right column as a dark shaded region.}
    \label{fig:eq_chem}
\end{figure}

\begin{table*}
    \caption{Retrieved parameters and their uncertainties.}
    \centering
    \renewcommand{\arraystretch}{1.5}
    \begin{tabular}{llllll}
    \hline
     Parameter                  & Description                               & Prior Range     & J1200                      & TWA28                      & J0856                      \\
    \hline
     $\log\ \mathrm{^{12}CO}$   & log mixing ratio of \twelveCO             & [-12.0, -2.0]   & $-4.0^{+0.1}_{-0.1}$       & $-3.7^{+0.1}_{-0.1}$       & $-3.7^{+0.1}_{-0.1}$       \\
     $\log\ \mathrm{^{13}CO}$   & log mixing ratio of \thirteenCO           & [-12.0, -2.0]   & $-6.1^{+0.2}_{-0.3}$       & $-5.6^{+0.2}_{-0.2}$       & $-5.6^{+0.1}_{-0.1}$       \\
     $\log\ \mathrm{H_2^{16}O}$      & log mixing ratio of $\mathrm{H_2^{16}O}$  & [-12.0, -2.0]   & $-4.2^{+0.1}_{-0.1}$       & $-3.9^{+0.1}_{-0.1}$       & $-3.8^{+0.1}_{-0.1}$       \\
     $\log\ \mathrm{H_2^{18}O}$ & log mixing ratio of \eighteenOwater       & [-12.0, -2.0]   & $-6.8^{+0.4}_{-1.6}$       & $-6.2^{+0.2}_{-0.3}$       & $-6.0^{+0.1}_{-0.1}$       \\
     $\log\ \mathrm{HF}$        & log mixing ratio of HF                    & [-12.0, -2.0]   & $-8.2^{+0.2}_{-0.2}$       & $-8.0^{+0.2}_{-0.2}$       & $-7.9^{+0.1}_{-0.2}$       \\
     $\log\ \mathrm{Na}$        & log mixing ratio of Na                    & [-12.0, -2.0]   & $-6.0^{+0.2}_{-0.2}$       & $-6.2^{+0.3}_{-0.3}$       & $-5.9^{+0.3}_{-0.5}$       \\
     $\log\ \mathrm{Ca}$        & log mixing ratio of Ca                    & [-12.0, -2.0]   & $-5.3^{+0.2}_{-0.2}$       & $-5.4^{+0.2}_{-0.2}$       & $-5.3^{+0.2}_{-0.2}$       \\
     $\log\ \mathrm{Ti}$        & log mixing ratio of Ti                    & [-12.0, -2.0]   & $-7.5^{+0.2}_{-0.3}$       & $-7.4^{+0.2}_{-0.2}$       & $-8^{+1}_{-2}$             \\
     $\log\ g$ [cm s$^{-2}$]                  & log surface gravity                       & [2.00, 5.50]    & $3.3^{+0.2}_{-0.2}$        & $3.5^{+0.1}_{-0.1}$        & $3.9^{+0.1}_{-0.1}$        \\
     $\epsilon_\mathrm{limb}$   & limb darkening coefficient                & [0.10, 0.98]    & $0.4^{+0.3}_{-0.2}$        & $0.2^{+0.1}_{-0.0}$        & $0.5^{+0.2}_{-0.2}$        \\
     $v\ \sin\ i$ [km s$^{-1}$]               & projected rotational velocity             & [2.0, 30.0]     & $5.3^{+0.2}_{-0.2}$        & $11.6^{+0.1}_{-0.1}$       & $4.5^{+0.2}_{-0.1}$        \\
     $v_\mathrm{rad}$ [km s$^{-1}$]           & radial velocity                           & [-40.00, 40.00] & $14.84^{+0.03}_{-0.04}$    & $12.6^{+0.1}_{-0.1}$       & $7.42^{+0.03}_{-0.03}$     \\
     $\nabla_{T,0}$             & temperature gradient at $P_0=10^2$ bar    & [0.06, 0.32]    & $0.2^{+0.1}_{-0.1}$        & $0.2^{+0.1}_{-0.1}$        & $0.2^{+0.1}_{-0.1}$        \\
     $\nabla_{T,1}$             & temperature gradient at $P_1+\Delta P$    & [0.06, 0.22]    & $0.14^{+0.04}_{-0.05}$     & $0.15^{+0.04}_{-0.05}$     & $0.14^{+0.05}_{-0.05}$     \\
     $\nabla_{T,2}$             & temperature gradient at $P_2+\Delta P$    & [0.06, 0.32]    & $0.09^{+0.04}_{-0.02}$     & $0.1^{+0.1}_{-0.0}$        & $0.09^{+0.03}_{-0.02}$     \\
     $\nabla_{T,3}$             & temperature gradient at $P_3+\Delta P$    & [0.06, 0.32]    & $0.063^{+0.004}_{-0.002}$  & $0.07^{+0.01}_{-0.00}$     & $0.063^{+0.003}_{-0.002}$  \\
     $\nabla_{T,4}$             & temperature gradient at $P_4+\Delta P$    & [0.06, 0.32]    & $0.15^{+0.02}_{-0.01}$     & $0.14^{+0.01}_{-0.01}$     & $0.16^{+0.01}_{-0.01}$     \\
     $\nabla_{T,5}$             & temperature gradient at $P_5+\Delta P$    & [0.02, 0.24]    & $0.03^{+0.01}_{-0.01}$     & $0.03^{+0.01}_{-0.01}$     & $0.03^{+0.01}_{-0.00}$     \\
     $\nabla_{T,6}$             & temperature gradient at $P_6+\Delta P$    & [0.00, 0.22]    & $0.01^{+0.01}_{-0.01}$     & $0.01^{+0.01}_{-0.01}$     & $0.01^{+0.01}_{-0.01}$     \\
     $\nabla_{T,7}$             & temperature gradient at $P_7=10^{-5}$ bar & [0.00, 0.22]    & $0.1^{+0.1}_{-0.1}$        & $0.1^{+0.1}_{-0.1}$        & $0.1^{+0.1}_{-0.1}$        \\
     $\log\Delta P$ [bar]            & log pressure shift of PT knots            & [-0.80, 0.80]   & $0.1^{+0.1}_{-0.1}$        & $0.3^{+0.1}_{-0.1}$        & $0.53^{+0.05}_{-0.05}$     \\
     $T_0$ [K]                      & temperature at $10^{2}$ bar               & [3000, 9000]    & $5002^{+627}_{-492}$       & $4930^{+626}_{-528}$       & $4086^{+368}_{-334}$       \\
     $r_0$                   & veiling factor at $\lambda = 1.90\mu$m    & [0.00, 2.00]    & $0.14^{+0.06}_{-0.06}$     & $0.02^{+0.02}_{-0.01}$     & $0.01^{+0.01}_{-0.01}$     \\
     $\alpha$                    & veiling power law exponent                & [0.00, 3.00]    & $0.74^{+0.58}_{-0.44}$     & $1.43^{+0.92}_{-0.87}$     & $1.34^{+0.89}_{-0.81}$     \\
     $\log\ a$                  & GP amplitude                              & [-1.00, 0.50]   & $0.282^{+0.005}_{-0.005}$  & $0.03^{+0.01}_{-0.01}$     & $0.333^{+0.004}_{-0.004}$  \\
     $\log\ l$ [nm]                  & GP lengthscale                            & [-2.00, -0.80]  & $-1.499^{+0.002}_{-0.014}$ & $-1.387^{+0.003}_{-0.004}$ & $-1.556^{+0.001}_{-0.007}$ \\
    \hline
    \end{tabular}
    \label{tab:retrieval_results}
    \tablefoot{The table includes the GP parameters, physical properties (surface gravity and rotational velocity), and VMRs of the chemical species. Uncertainties are represented as $1\sigma$ intervals.}
    \end{table*}
In this work we adopted an approach where the mixing ratios of line species were treated as independent parameters, a method we refer to as `free composition'. This approach does not enforce constraints on the relative abundances, assuming constant-with-altitude mixing ratios. Given the limited vertical extent probed by the observed spectral lines, this assumption is generally valid for many scenarios. Nonetheless, it is insightful to contrast our retrieved abundances against those predicted by EC.
The EC profiles for most species display minimal variation across the temperature and pressure ranges of our subjects, except for titanium (Ti), which shows a notable decrease in concentration with rising temperatures (Fig. \ref{fig:eq_chem}). Using \texttt{FastChem}\footnote{\url{https://github.com/exoclime/fastchem}} \citep{kitzmannFastchemCondEquilibrium2023}, we compared our retrieved mixing ratios against EC predictions, considering elemental solar abundances adjusted of the retrieved metallicity and scaling the abundance of \twelveCO{} to match the retrieved C/O. To adjust for any potential systematic offsets due to the metallicity-surface gravity degeneracy, we normalised the VMRs relative to VMR($^{12}$CO).

Our findings indicate that the vertical distributions of the primary molecular species ($^{12}$CO and H$_2$O) are relatively stable over the pressure-temperature range of the three objects. In general, the retrieved values align with the EC predictions to within a $1\sigma$ confidence level, with the significant exception of calcium (Ca). While EC predicts comparable mixing ratios for sodium (Na) and Ca, our analysis find the retrieved VMRs of Na  to be consistent with EC, while the Ca abundances are retrieved at significantly higher values than EC. However, the contribution of Ca mostly originates from a small region at the bluest order of \CRIRES (see Fig. \ref{fig:opacity_sources}), where telluric lines are prevalent (see Fig. \ref{fig:telluric_correction}); hence, the interpretation of the Ca abundance is not straightforward.
For Ti, the retrieved mixing ratios broadly align to EC in the atmosphere's deepest regions while for HF the retrieved mixing ratios are in the range of $10^{-8}$ to $10^{-7}$,  resulting in close agreement with EC.  

\subsection{Isotopic composition}\label{subsec:isotopic_composition}
\subsubsection{Carbon isotope}
The obtained \Cratio\ for J1200, TWA 28, and J0856 (see Fig. \ref{fig:bestfit_chemistry}),

\begin{align*}
    \mathrm{\textsuperscript{12}C/\textsuperscript{13}C}_\text{ J1200} & = 114^{+69}_{-33},\\ 
    \mathrm{\textsuperscript{12}C/\textsuperscript{13}C}_\text{ TWA28} & = 81^{+28}_{-19},\\ 
    \mathrm{\textsuperscript{12}C/\textsuperscript{13}C}_\text{ J0856} & = 79^{+20}_{-14},\\ 
\end{align*}
are consistent among the three objects within 1$\sigma$ uncertainties. The retrieved values are in agreement with the ISM given the scatter around the present-day value \Cratioism = $68 \pm 16$ \citep{milam1213Isotope2005} and show no significant enrichment in \thirteenCO. In the context of young substellar objects, our results are in good agreement with other studies that have reported \Cratio\ values consistent with the ISM \citep{xuanValidationElementalIsotopic2023,gandhiJWSTMeasurements13C2023a}. The retrieved 1$\sigma$ uncertainties cover the solar value \Cratiosolar $\sim$ 89 \citep{meijaIsotopicCompositionsElements2016} and partially overlap with objects of higher \Cratio, $\sim 100$ \citep{zhang12CO132021,costesFreshViewHot2024, hoodHighPrecisionAtmosphericConstraints2024}. In contrast to the first object of the SupJup Survey: DENIS J0255 (\Cratio=$184^{+61}_{-40}$; \cite{deregtESOSupJupSurvey2024a}), the retrieved \Cratio\ values are consistent with the ISM, indicating no significant enrichment or depletion of \thirteenCO{} in the atmospheres of these objects. As noted in \cite{deregtESOSupJupSurvey2024a}, the age of the objects may play a significant role in the isotopic composition, with younger objects exhibiting a lower \Cratio\ due to the \thirteenCO{} enrichment in the ISM over time \citep{milam1213Isotope2005,romanoEvolutionCNOIsotopes2017}.

\subsubsection{Oxygen isotope}
The detection of \eighteenOwater{} in TWA 28 and J0856, and the marginal detection in J1200, provides additional insights into the isotopic composition of these objects. From the retrieved \sixteenOwater{} and \eighteenOwater{} VMRs, we calculated the \ratio{O}{16}{18} ratio for each object,
\begin{align*}
    \mathrm{\textsuperscript{16}O/\textsuperscript{18}O}_\text{ J1200} & > 159,\\ 
    \mathrm{\textsuperscript{16}O/\textsuperscript{18}O}_\text{ TWA28} & = 205^{+140}_{-62},\\ 
    \mathrm{\textsuperscript{16}O/\textsuperscript{18}O}_\text{ J0856} & = 141^{+42}_{-28},\\ 
\end{align*}
where the lower limit for J1200 is due to its marginal detection. The retrieved \ratio{O}{16}{18} ratios are lower than the solar value of $525\pm 21$ \citep{lyonsLightCarbonIsotope2018} and the ISM value of $557\pm 30$ \citep{wilsonIsotopesInterstellarMedium1999}, indicating a potential enrichment of \eighteenO{} in the atmospheres of these objects. The \ratio{O}{16}{18} ratios reported for HIP 55507 B (\ratio{O}{16}{18}=$288^{+125}_{-70}$; \cite{xuanValidationElementalIsotopic2023}) also exhibited a lower value than the ISM (i.e. a higher \eighteenO{} abundance). \cite{gandhiJWSTMeasurements13C2023a} showcased the excellent capabilities of JWST to measure the \ratio{O}{16}{18} ratio in the atmospheres of young substellar objects, with a reported value of \ratio{O}{16}{18}=$425^{+33}_{-28}$ for the young SJ VHS 1256 b. The \ratio{O}{16}{18} retrieved in this work for J0856, with the caveat that \eighteenOwater{} is only detected at 3.3$\sigma$, is one of the lowest values reported for young substellar objects, indicating a potential deviation of \eighteenO{} in the atmosphere of this object from the ISM value.

\subsection{Surface gravity}\label{subsec:logg_metallicity}
Surface gravity significantly influences the observed spectra by affecting line shapes, the temperature profile, and the overall continuum \citep{marleyCoolSideModeling2015,mukherjeeSonoraSubstellarAtmosphere2024}. Its inverse relationship with atmospheric scale height impacts the pressure range of the photosphere, a key property determining spectral features. We report low surface gravities for the three objects (see \Cref{tab:retrieval_results}), consistent with expectations for young objects \citep{baraffeEvolutionaryModelsLowmass2002,allersCharacterizingYoungBrown2007,bonnefoyLibraryNearinfraredIntegral2014}.

\begin{figure}
    \centering
    \includegraphics[width=\linewidth]{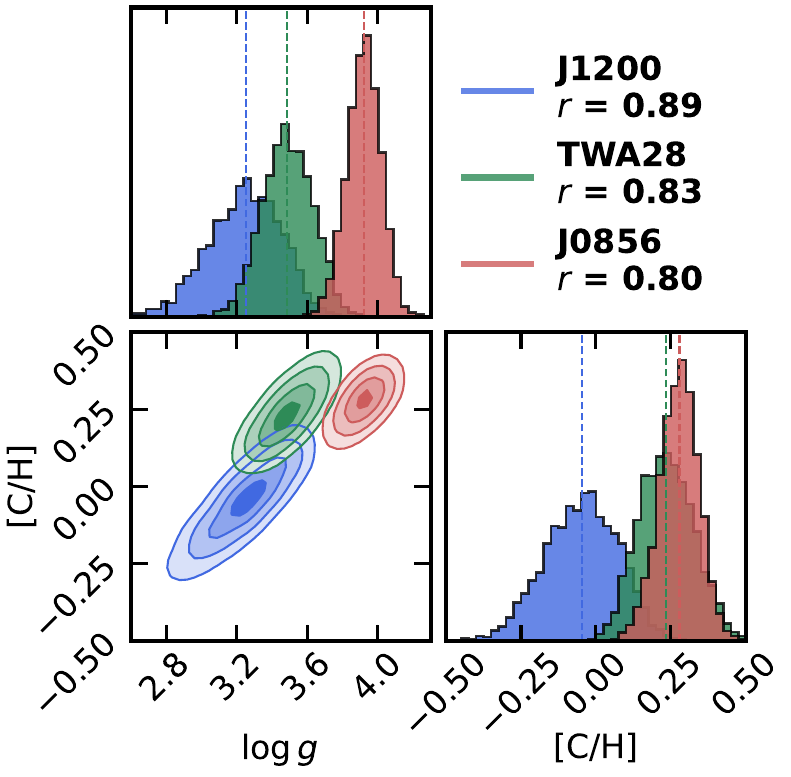}
    \caption{Posterior distributions of the surface gravity and carbon-to-oxygen ratio relative to the solar value for each object. The contours of the 2D histogram represent the 1, 2, and 3$\sigma$ confidence intervals, with dashed lines indicating the median values. The Pearson correlation coefficient is shown in the top-right corner, where a value of $r=0$ indicates no correlation and $r=1$ indicates a perfect positive correlation.}
    \label{fig:bestfit_logg_FeH}
\end{figure}

The posterior distributions are highly correlated\footnote{Correlation is quantified by the Pearson coefficient, calculated using \href{https://docs.scipy.org/doc/scipy/reference/generated/scipy.stats.pearsonr.html}{scipy.stats.pearsonr}, where a value of $r=1$ indicates perfect positive correlation and $r=0$ indicates no correlation.} (see Fig. \ref{fig:bestfit_logg_FeH}), making it difficult to precisely constrain surface gravities and metallicities. Determining surface gravity from \textit{K}-band observations is challenging due to the scarcity of reliable gravity-sensitive features \citep{zhang12CO132021}. This limitation could be mitigated with additional data at shorter (e.g. J- or H-band) wavelengths. Furthermore, incorporating prior knowledge of precise dynamical mass and radius measurements could help resolve existing degeneracies \citep{stolkerMIRACLESAtmosphericCharacterization2020}.

\subsection{Veiling}

The best-fit models for TWA 28 and J0856 show no evidence of veiling in their spectra, that is, $r_0 \approx 0.0$. However, for J1200, the retrieved veiling factor is $r_0 = 0.14^{+0.06}_{-0.06}$, suggesting the presence of a weak veiling continuum. The wavelength dependence of the veiling factor is loosely constrained by the data, favouring a power-law index of $\alpha = 0.74^{+0.58}_{-0.44}$, which increases towards redder wavelengths.

Veiling is a widely observed phenomenon in the near-infrared spectra of many young stellar objects, particularly accreting sources \citep{folhaHighVeilingInfrared1999,mcclureCHARACTERIZINGSTELLARPHOTOSPHERES2013,sullivanOpticalNearinfraredExcesses2022,sousaNewInsightsNearinfrared2023}. However, it is less common in substellar objects, with only a few detections in young BDs (e.g. \citealt{whiteVeryLowMass2003}). The physical origin of veiling is still debated, with potential sources including accretion, chromospheric activity, and magnetic fields \citep{hartiganHowUnveilTauri1989,folhaHighVeilingInfrared1999,stempelsPhotosphereVeilingSpectrum2003,reiLinedependentVeilingVery2018}.

There is a potential link between mass accretion rate and near-infrared veiling; more massive objects with higher accretion rates are expected to exhibit stronger veiling \citep{whiteVeryLowMass2003,sousaNewInsightsNearinfrared2023}. The presence of veiling in J1200 may indicate ongoing accretion, which is plausible given its young age, the detection of a disk around this object \citep{schutteDiscoveryNearbyYoung2020}, and its higher mass compared to TWA 28 and J0856 (see \Cref{tab:fundamental_parameters_reshaped}).

\subsection{Correlated noise}
Our analysis revealed significant correlation in the data as reflected by the retrieved parameters of the GP kernel. We observe a tentative relation between the GP amplitude, \(a\), and length scale, \(l\), and the spin of the objects. Specifically, slow rotators, characterised by narrower spectral features, exhibit a smaller GP length scale and a larger amplitude compared to fast rotators. The retrieved values, ordered from the slowest to the fastest rotator, are presented below:

\begin{align*}
    \text{J0856}& \quad (v\sin i \approx 4.5 \text{ km s}^{-1})\\ 
    a & = 2.15^{+0.02}_{-0.02} \\ 
    l & = 0.0278^{+0.0001}_{-0.0004}\approx 3.6 \text{ pixels} \\ 
    \text{J1200}& \quad (v\sin i \approx 5.3 \text{ km s}^{-1})\\ 
    a & = 1.91^{+0.02}_{-0.02} \\ 
    l & = 0.0317^{+0.0002}_{-0.0010}\approx 4.1 \text{ pixels} \\ 
    \text{TWA28}& \quad (v\sin i \approx 11.6 \text{ km s}^{-1})\\ 
    a & = 1.06^{+0.01}_{-0.01} \\ 
    l & = 0.0410^{+0.0002}_{-0.0003}\approx 5.3 \text{ pixels.} \\ 
\end{align*}
In this context, the GP amplitude serves as a dimensionless parameter adjusting the covariance matrix's off-diagonal elements, while the length scale, expressed in nanometers (nm), reflects the correlation distance over which the correlation between two points decreases to \(1/e\) of its initial value. This scale generally matches the width of the spectral lines within the observed spectra. Employing the \CRIRES sampling rate, the length scale is converted to a number of pixels. Accurately accounting for correlated noise is essential for obtaining reliable uncertainties in the retrieved parameters. We performed a test retrieval without the GP kernel to assess the impact of correlated noise on the retrieved parameters. The results show that the uncertainties in the retrieved parameters are underestimated when correlated noise is not accounted for (see Fig. \ref{fig:cornerplot_J0856_GP}).
 This consideration is particularly pertinent to HRS, where correlated noise can substantially alter the interpretation of the retrieved parameters \citep{deregtESOSupJupSurvey2024a}.

\section{Conclusions}\label{sec:conclusions}

We analysed high-resolution CRIRES$^+$ data for three young BDs, gaining insights into their atmospheric composition, including carbon and oxygen isotopes, thermal structure, rotational velocities, and the presence of veiling. The retrieved thermal profiles of the three objects have similar structures, reflecting their comparable SpTs. The surface gravities and rotational velocities are consistent with those of young, low-gravity objects, indicating the slow rotation typical of young BDs still undergoing contraction.

The retrieved metallicities are broadly consistent with solar values, but their interpretation is complicated by the strong correlation between surface gravity and metallicity. The uncertainties in surface gravity are likely underestimated due to this correlation (see Fig. \ref{fig:bestfit_logg_FeH}). We improved the existing methodology for atmospheric retrievals by incorporating correlated noise, which is crucial for HRS (see \Cref{app:correlated_noise_results}). Our results highlight the importance of understanding and accounting for these correlations when interpreting atmospheric parameters, especially in the context of degeneracies between surface gravity and metallicity.

Precise measurements of the \Cratio\ isotope ratios are challenging, yet we demonstrate that HRS can constrain the isotopic composition of BDs. The measured \Cratio\ in our sample are consistent with that of the ISM within 1$\sigma$, considering the scatter of the ISM value \Cratioism = $68\pm15$. This supports the idea that isolated BDs share a common formation mechanism with stars, likely through fragmentation processes. These findings add valuable data points for assessing the role of the \Cratio\ as a formation tracer and suggest that oxygen isotopes might be accessible with current and future \textit{K}-band HRS. The measured C/Os are consistent with the solar value and show no significant enrichment in carbon or oxygen. Additional measurements of the C/O in young BDs are needed to assess the role of this parameter in the formation and evolution of substellar objects.
Looking ahead, expanding our sample size to include more isolated BDs and substellar companions, as planned in the ESO SupJup Survey,  will enhance our understanding of the \Cratio\ as a potential formation tracer and deepen our insights into the diverse atmospheres of BDs and SJs, including new measurements of the C/O across objects of different ages and SpTs. Additional isotope ratios might be within the reach of state-of-the-art high-resolution spectrographs, such as CRIRES$^+$, and current space-based facilities, such as JWST (e.g. \citealt{gandhiJWSTMeasurements13C2023a,barrado15NH3AtmosphereCool2023}). In the future, high-resolution spectrographs on next-generation ground-based telescopes, such as the METIS instrument at the E-ELT \citep{brandlMETISMidinfraredELT2021}, will be powerful tools for characterising the atmospheres and measuring isotope ratios of cool BDs and directly imaged exoplanets.

\begin{acknowledgements}
D.G.P and I.S. acknowledge NWO grant OCENW.M.21.010. Based on observations collected at the European Organisation for Astronomical Research in the Southern Hemisphere under ESO programme(s) 1110.C-4264(F). This work used the Dutch national e-infrastructure with the support of the SURF Cooperative using grant no. EINF-4556. This research has made use of NASA's Astrophysics Data System. This research has made use of adstex (\url{https://github.com/yymao/adstex}).
\newline
\textit{Software}: \texttt{NumPy} \citep{harrisArrayProgrammingNumPy2020}, \texttt{SciPy} \citep{virtanenSciPyFundamentalAlgorithms2020}, \texttt{Matplotlib} \citep{hunterMatplotlib2DGraphics2007}, \texttt{petitRADTRANS} \citep{mollierePetitRADTRANSPythonRadiative2019}, \texttt{PyAstronomy} \citep{czeslaPyAPythonAstronomyrelated2019}, \texttt{Astropy} \citep{collaborationAstropyProjectSustaining2022}, \texttt{corner} \citep{foreman-mackeyCornerPyScatterplot2016}.
\end{acknowledgements}

\bibliographystyle{aa_url}
\bibliography{adsbib}

\onecolumn
\appendix

\section{Telluric correction with \texttt{Molecfit}}\label{app:molecfit}
\begin{figure}[h!]
    \centering
    \includegraphics[width=\linewidth]{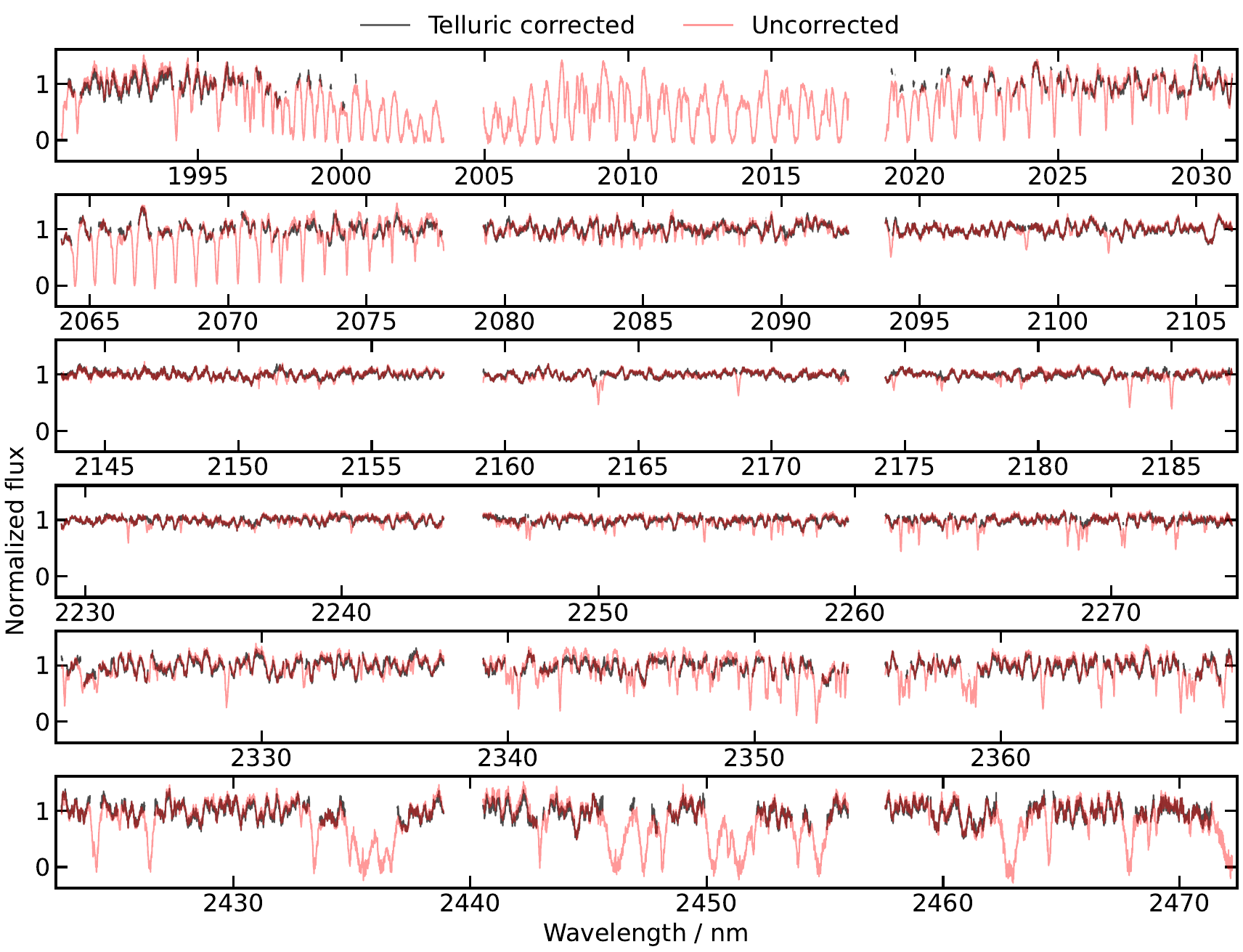}
    \caption{Telluric correction for TWA 28. Each panel shows a different spectral order. The corrected spectrum (black) is the result of dividing the data by the telluric model (red). Telluric lines deeper than 0.65 (continuum normalised to 1.0) with respect to the continuum are masked for the atmospheric retrieval.}
    \label{fig:telluric_correction}
\end{figure}

The open-source software \texttt{Molecfit} \citep{smetteMolecfitGeneralTool2015} is a tool for correcting astronomical observations for the transmission of the Earth's atmosphere. \texttt{Molecfit} fits a telluric model to the observed data based on airmass, precipitable water vapour, the thermal profile of the Earth's atmosphere at the time of the observations and the abundances of the main opacity sources (H2O, CO2, CO, CH4, and N2O). During the fitting process, the wavelength solution and the continuum of the model are adjusted to match the observed spectrum. Additionally, \texttt{Molecfit} fits for the instrumental profile of the spectrograph, which is crucial for HRS. The best-fit telluric model from the standard star is then scaled to the airmass and precipitable water vapour of the science observations. Lastly, the telluric model is divided from the observed data to obtain the telluric-corrected spectrum (see Fig. \ref{fig:telluric_correction}). The refined wavelength solution and the continuum of the telluric model are used to calibrate the data. The fitted continuum on the standard star is divided by the black body function corresponding to the effective temperature of the star to obtain the wavelength-dependent throughput. From the fitting of the line spread function with a Lorentzian profile we obtain an empirical estimate of the spectral resolution of the instrument,
\begin{align}
    R = \frac{\lambda}{\Delta \lambda} = \frac{c}{\Delta v}=\frac{c}{\text{FWHM}} = 50,000 \pm 1,500,
\end{align}
which is in perfect agreement with the nominal resolution of CRIRES$^+$ in the wide slit mode ($R\sim 50,000$) and is equivalent to a resolution element of $\Delta v \sim 3$ km s$^{-1}$.

\newpage
\section{Hydrogen fluoride}\label{app:HF}
\begin{figure}[h]
    \centering
    \includegraphics[width=\linewidth]{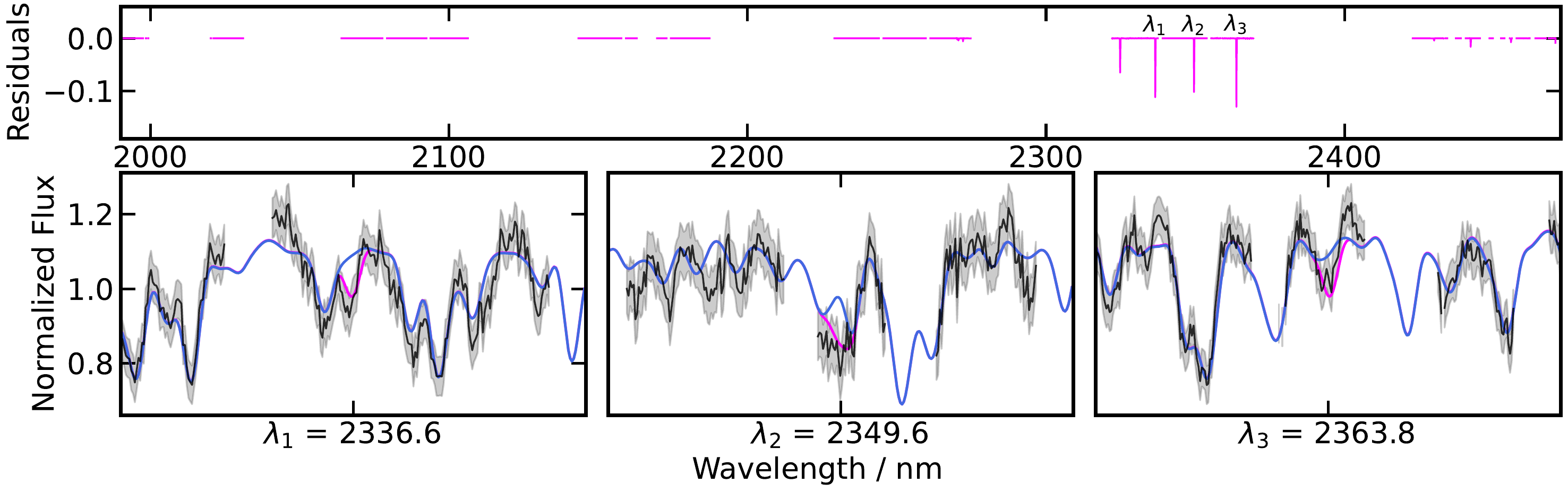}
    \caption{Top: Retrieved HF line depths for J1200. Bottom: Best-fitting models with HF (magenta) and without HF (blue), plotted around three regions with strong HF lines.}
    \label{fig:HF_spec_J1200}
\end{figure}
\begin{figure}[h]
    \centering
    \includegraphics[width=\linewidth]{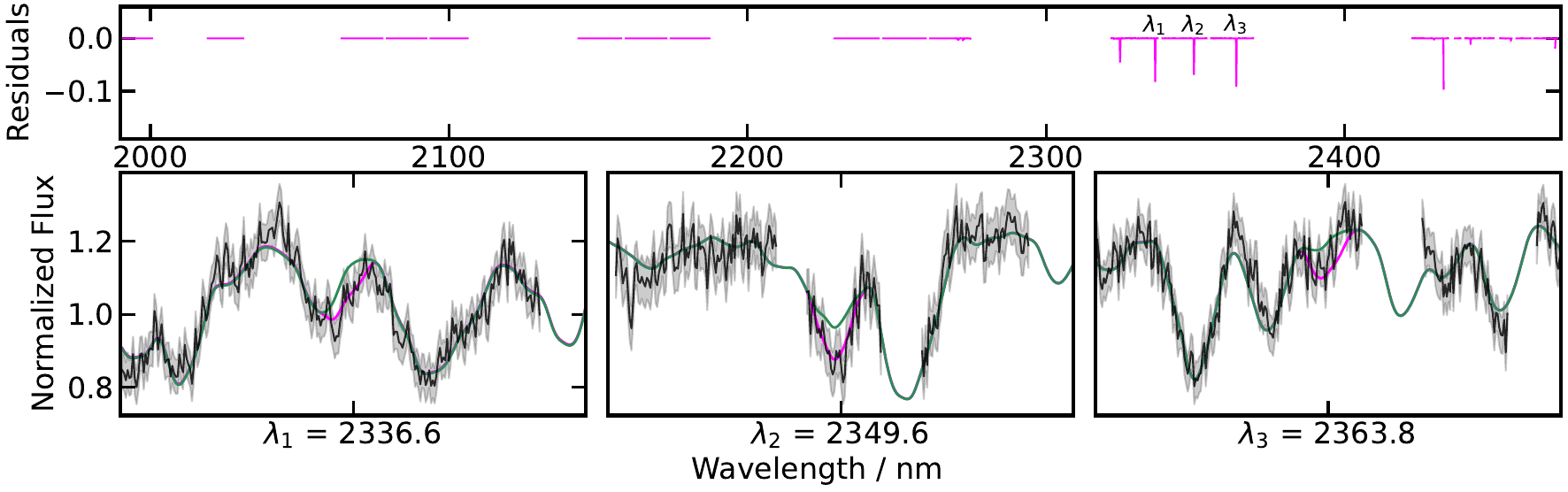}
    \caption{Top: Retrieved HF line depths for TWA 28. Bottom: Best-fitting models with HF (magenta) and without HF (green), plotted around three regions with strong HF lines.}
    \label{fig:HF_spec_TWA28}
\end{figure}
\begin{figure}[h]
    \centering
    \includegraphics[width=\linewidth]{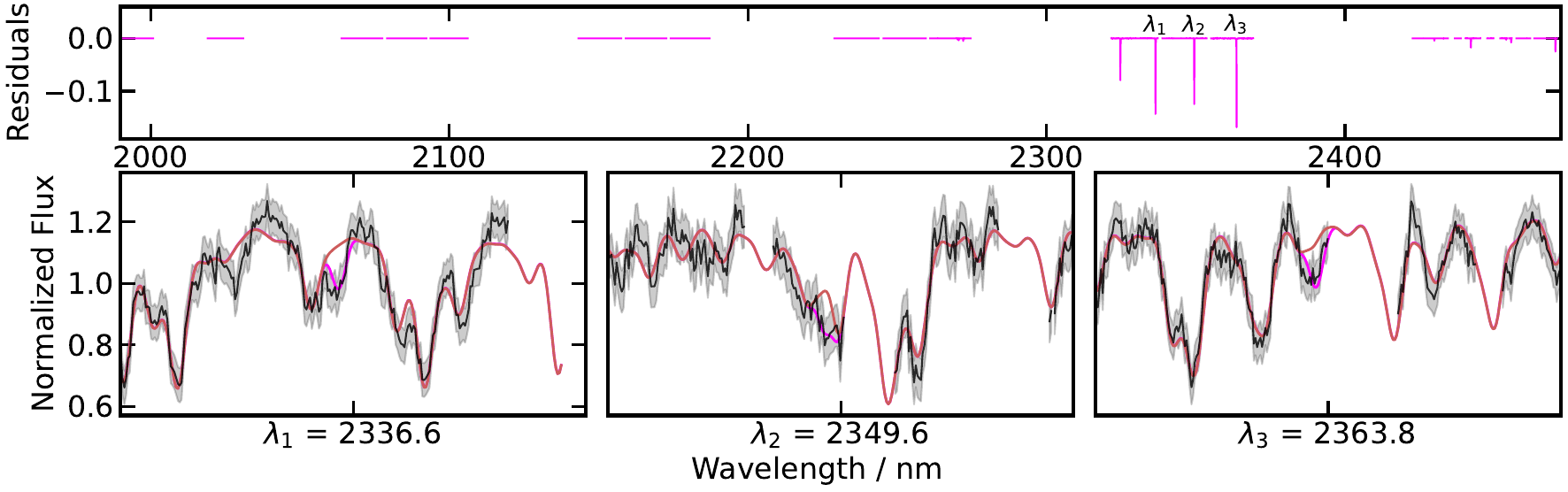}
    \caption{Top: Retrieved HF line depths for J0856. Bottom: Best-fitting models with HF (magenta) and without HF (red), plotted around three regions with strong HF lines.}
    \label{fig:HF_spec_J0856}
\end{figure}

\newpage
\section{Extended corner plots}\label{app:cornerplot}
\begin{figure}[ht]
    \centering
    \includegraphics[width=\linewidth]{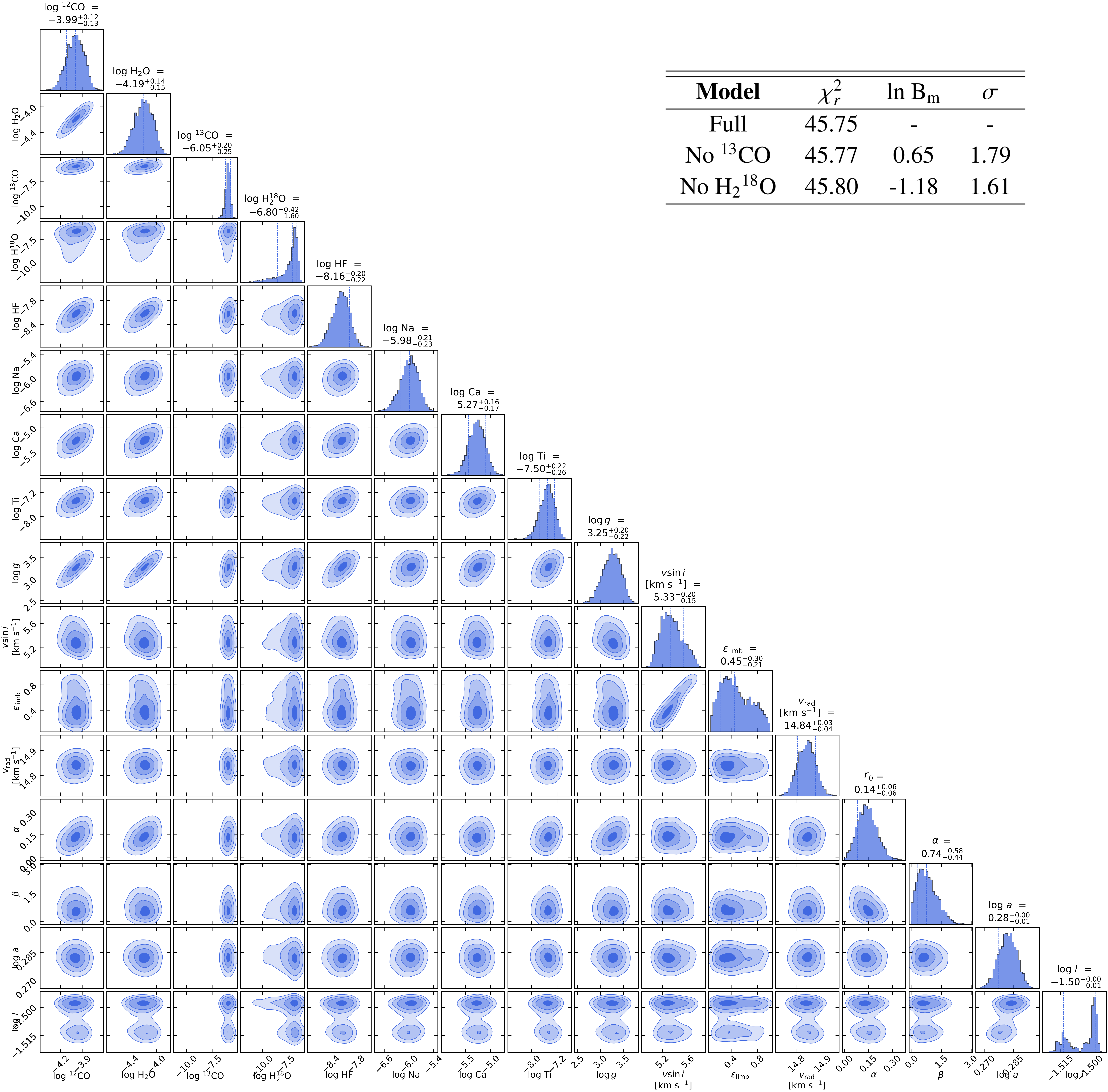}
    \caption{Posterior distributions of the retrieved parameters of J1200. The reduced chi-squared values for the models with and without \thirteenCO{} and \eighteenOwater{} are shown in the top-right corner along with the Bayes factors, $B_{\rm m}$, and the corresponding significance levels.}
    \label{fig:cornerplot_J1200}
\end{figure}
\begin{figure}[ht]
    \centering
    \includegraphics[width=\linewidth]{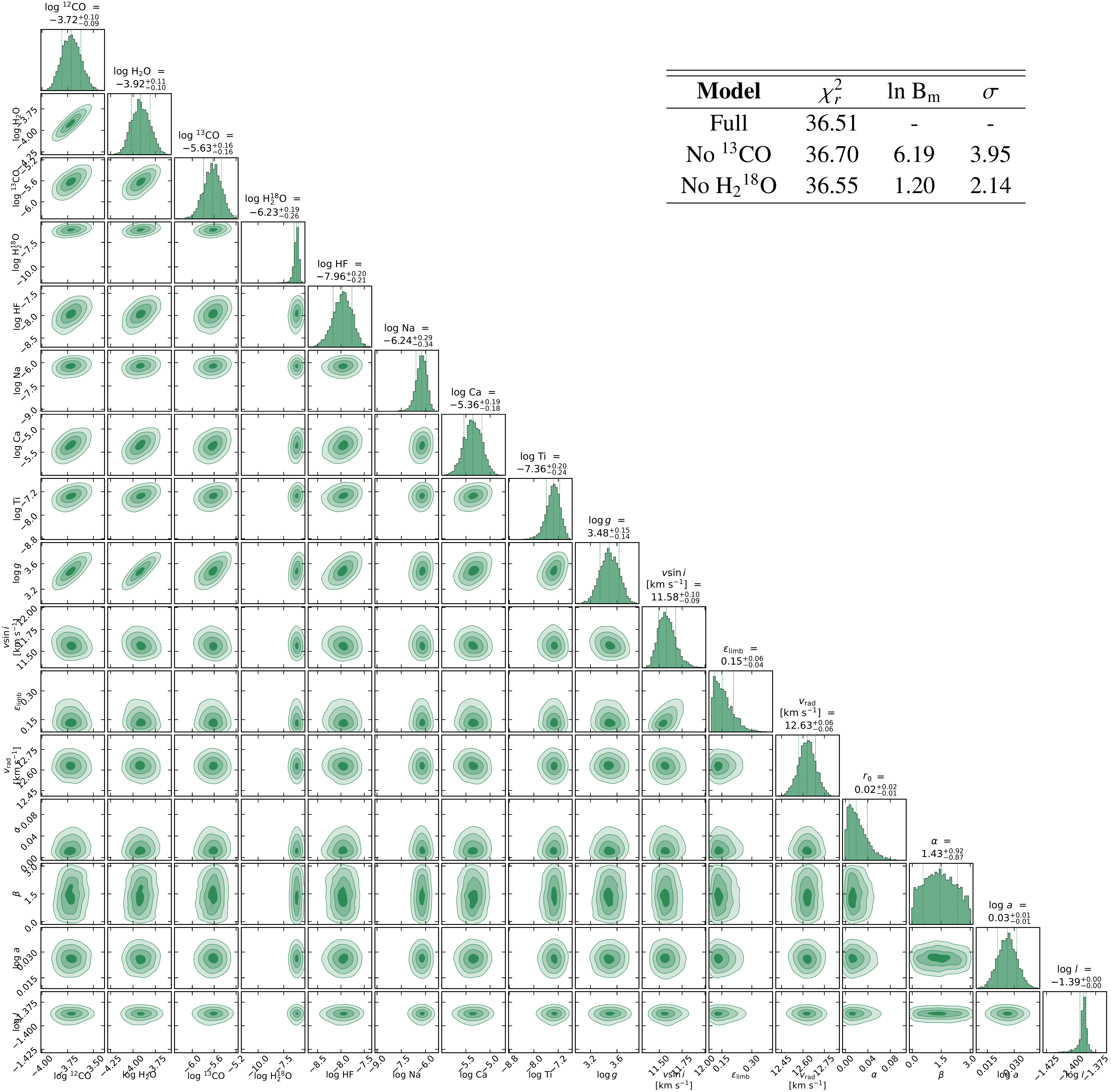}
    \caption{Posterior distributions of the retrieved parameters of TWA 28. The reduced chi-squared values for the models with and without \thirteenCO{} and \eighteenOwater{} are shown in the top-right corner along with the Bayes factors, $B_{\rm m}$, and the corresponding significance levels.}
    \label{fig:cornerplot_TWA28}
\end{figure}

\begin{figure}[ht]
    \centering
    \includegraphics[width=\linewidth]{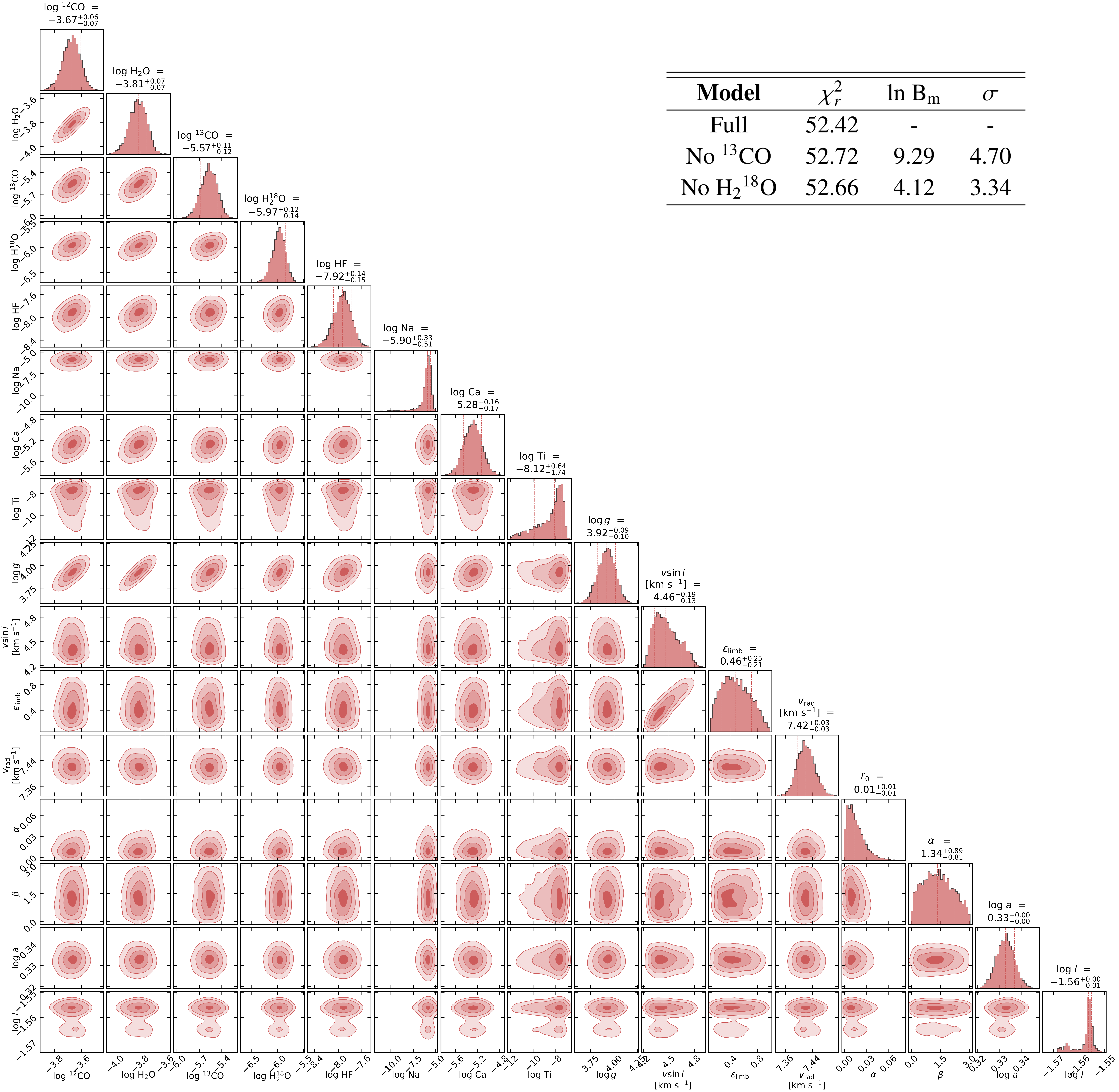}
    \caption{Posterior distributions of the retrieved parameters of J0856. The reduced chi-squared values for the models with and without \thirteenCO{} and \eighteenOwater{} are shown in the top-right corner along with the Bayes factors, $B_{\rm m}$, and the corresponding significance levels.}
    \label{fig:cornerplot_J0856}
\end{figure}

\clearpage
\section{Best-fitting spectra}\label{app:bestfit_spectra}
\begin{figure*}[h]
    \centering
    \includegraphics[width=\linewidth]{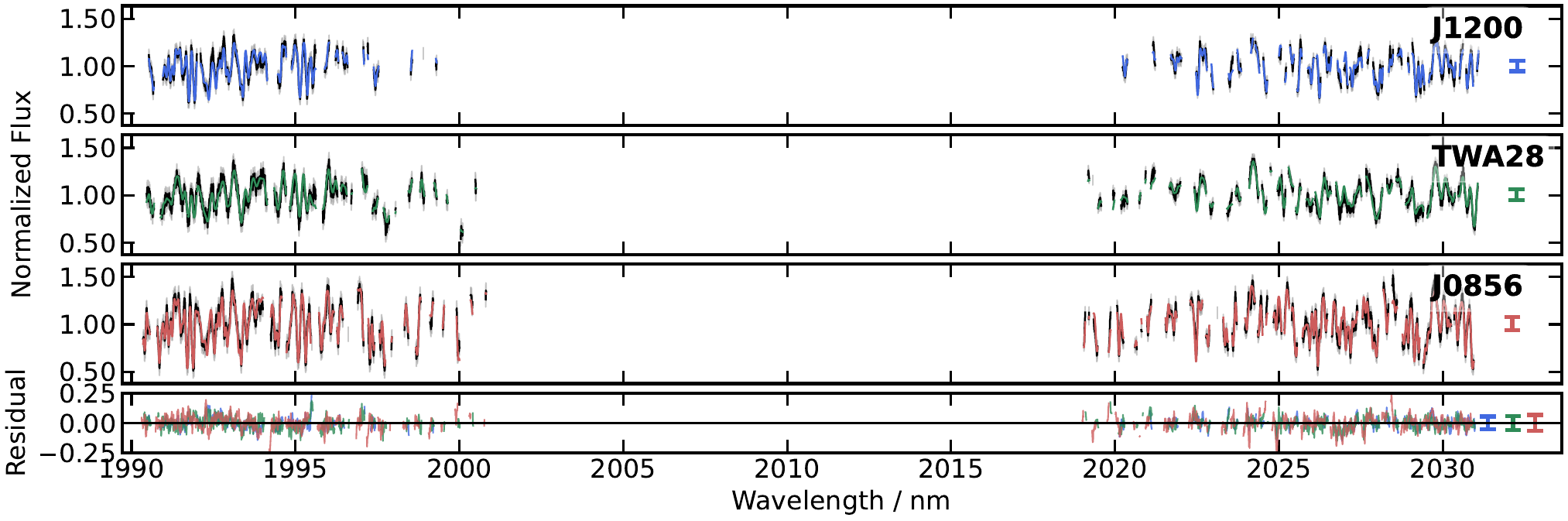}
    \caption{The best-fitting models of the three objects for the first (bluest) spectral order. The central detector is masked due to the presence of strong telluric lines.}
    \label{fig:bestfit_spectra_order0}
\end{figure*}

\begin{figure*}[h]
    \centering
    \includegraphics[width=\linewidth]{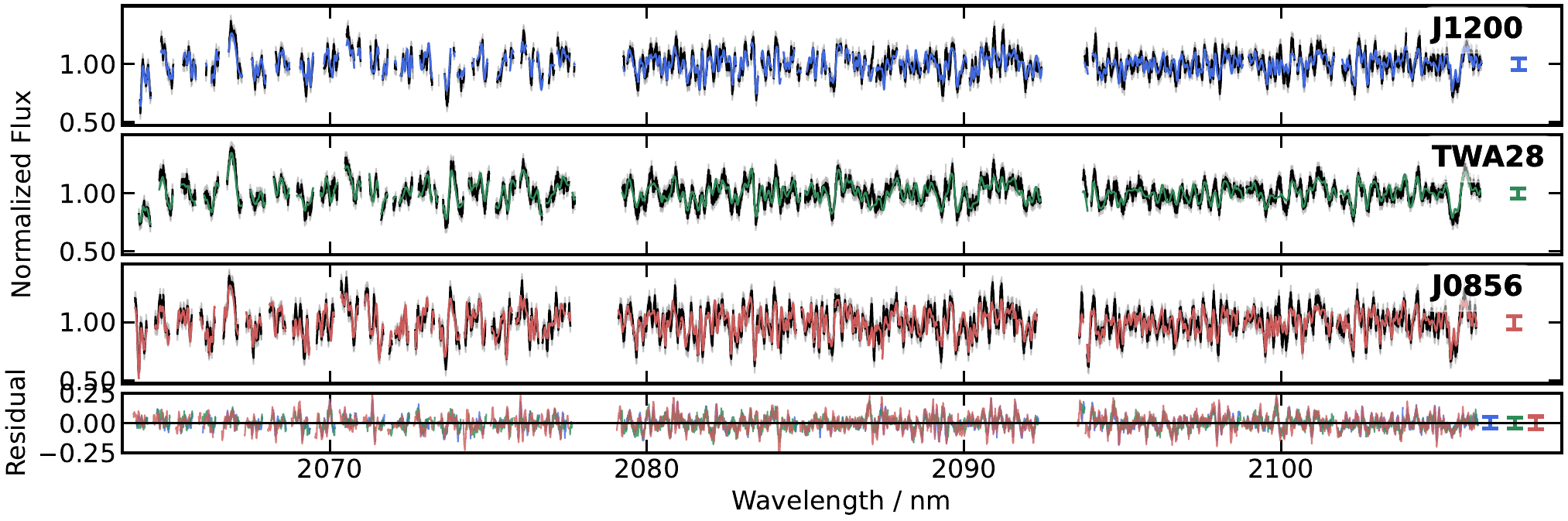}
    \caption{The best-fitting models of the three objects for the second spectral order.}
    \label{fig:bestfit_spectra_order1}
\end{figure*}

\begin{figure*}[h]
    \centering
    \includegraphics[width=\linewidth]{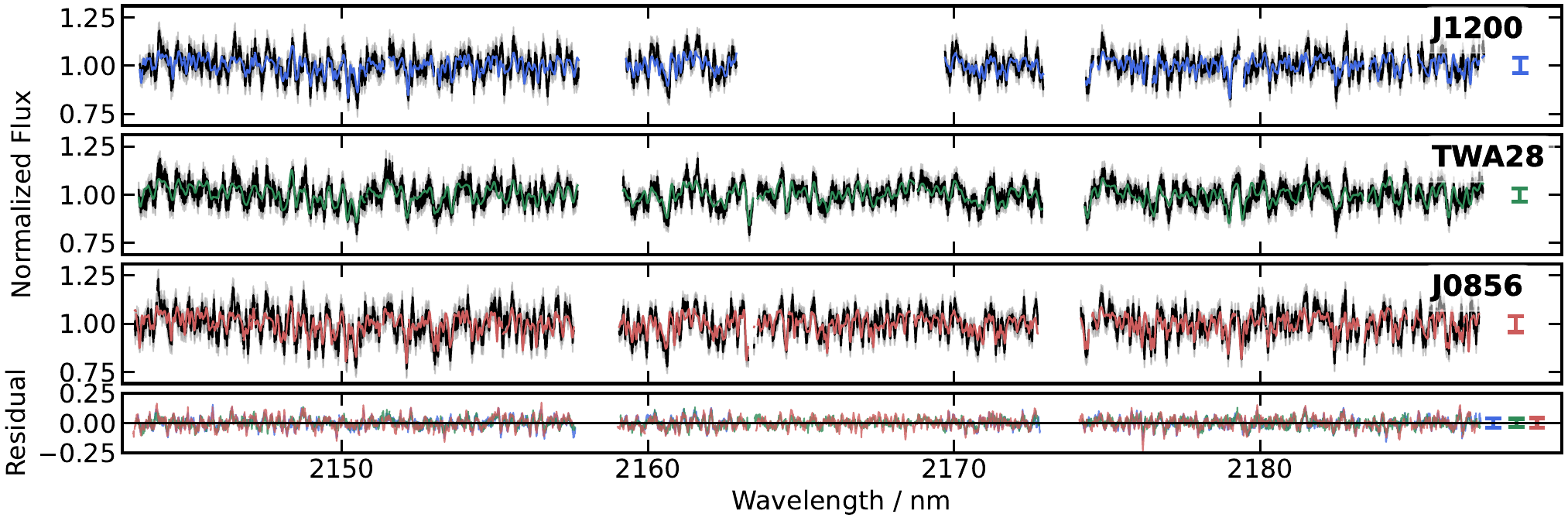}
    \caption{The best-fitting models of the three objects for the third spectral order. The Brackett-$\gamma$ line  ($\lambda$2166 nm) of the standard star affects the central detector of this spectral order for J1200. It is masked at the preprocessing stage.}
    \label{fig:bestfit_spectra_order2}
\end{figure*}

\begin{figure*}[h!]
    \centering
    \includegraphics[width=\linewidth]{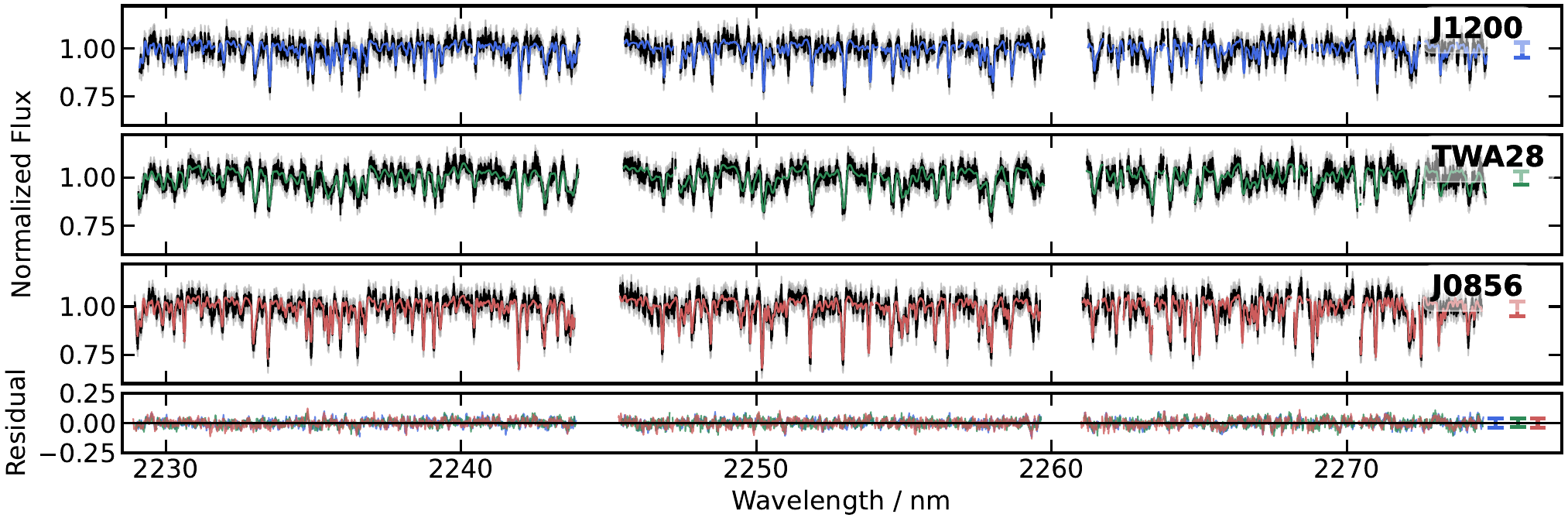}
    \caption{The best-fitting models of the three objects for the fourth spectral order.}
    \label{fig:bestfit_spectra_order3}
\end{figure*}

\begin{figure*}[h!]
    \centering
    \includegraphics[width=\linewidth]{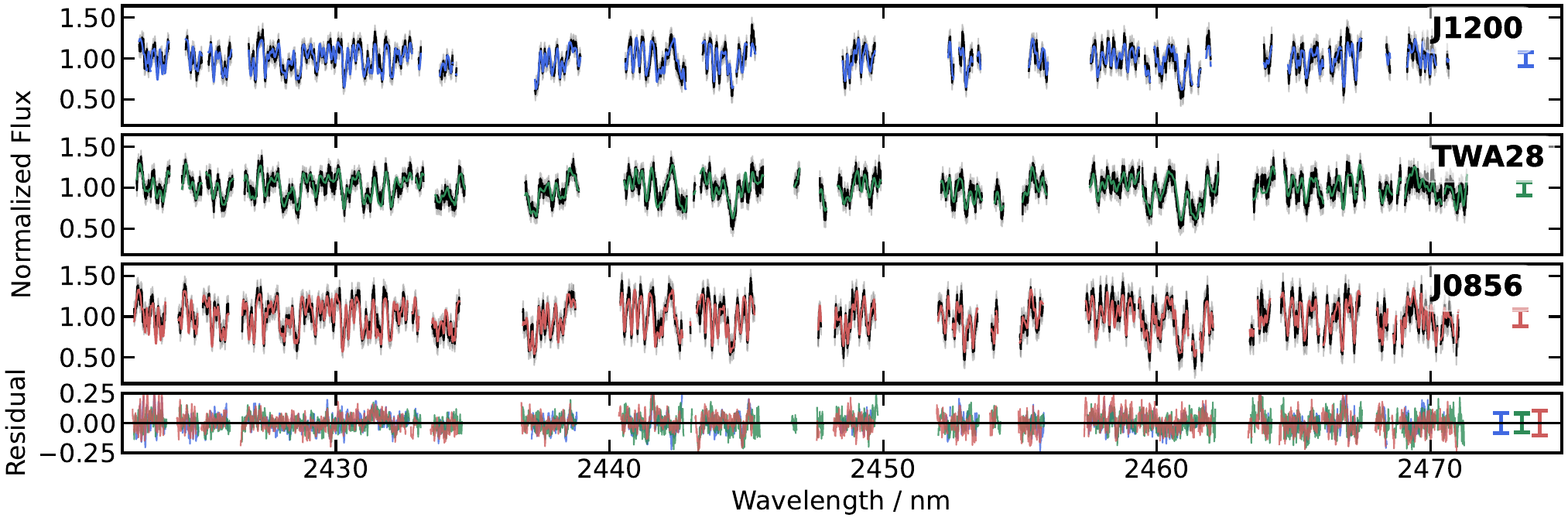}
    \caption{The best-fitting models of the three objects for the sixth spectral order. This region contains several strong telluric lines that are masked at the preprocessing stage.}
    \label{fig:bestfit_spectra_order5}
\end{figure*}

\clearpage
\section{Correlated noise comparison}\label{app:correlated_noise_results}
\begin{figure}[h]
    \centering
    \includegraphics[width=\linewidth]{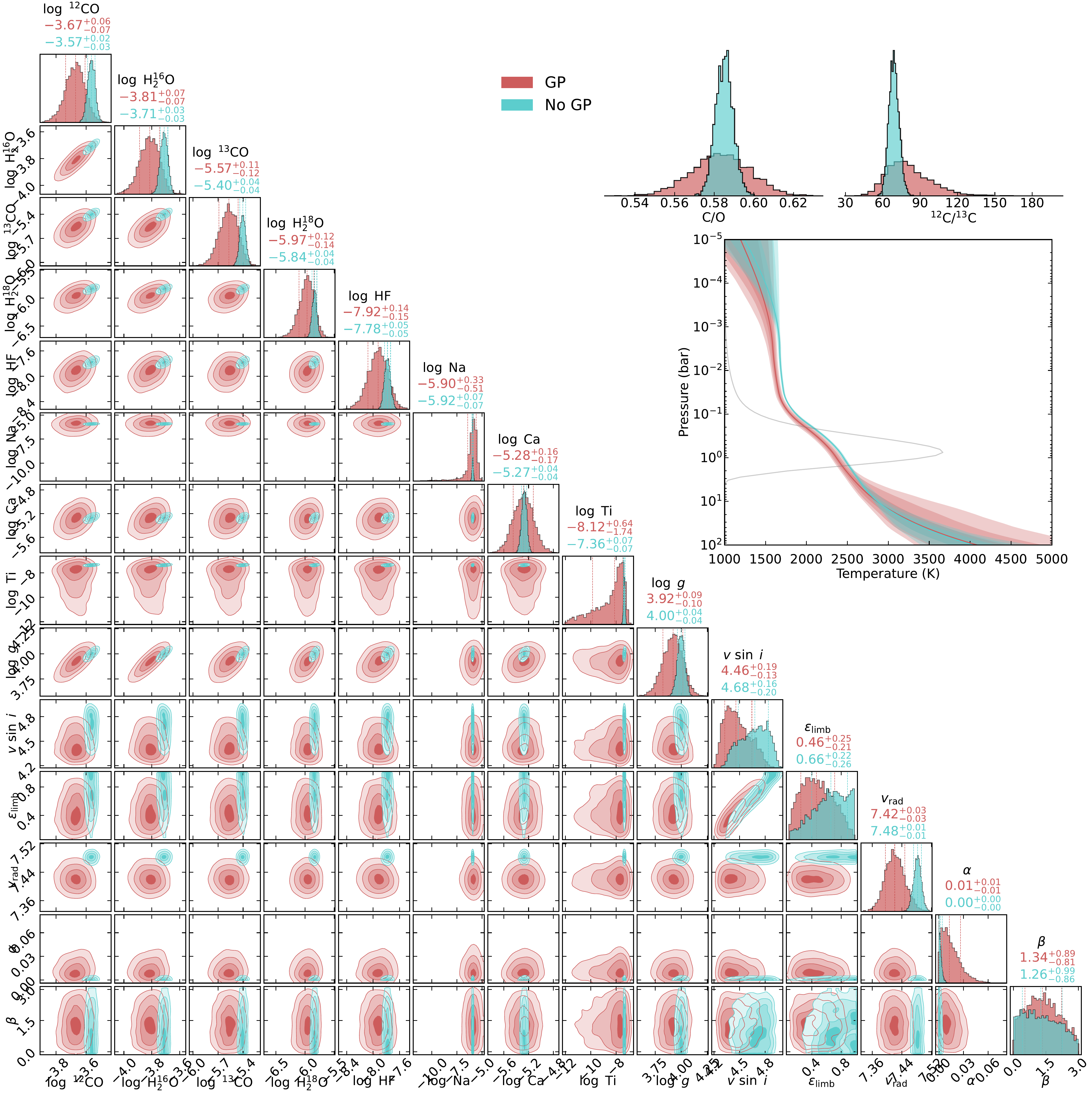}
    \caption{Posterior distributions of the retrieved parameters of J0856: including correlated noise with a GP (red) and without accounting for correlated noise (cyan).}
    \label{fig:cornerplot_J0856_GP}
\end{figure}
\end{document}